%Paper: hep-th/9310052
%From: CASSANDRO@roma1.infn.it
%Date: Fri, 8 Oct 1993 10:44:11 +0100 (WET)

\magnification\magstep1
\hsize=15.5truecm
\vsize=25truecm
\baselineskip=12pt
\tolerance=1000
%\nopagenumbers
%
%  M A C R O S
%
% title author and address
%
\long\def\title#1{\parindent 0pt
{\baselineskip 24pt\tolerance=10000
\vglue 4.2truecm
\noindent{#1}}\par}
\long\def\author#1{\vskip 12pt\begingroup{
\tolerance=10000\parindent=3truecm #1}\par\endgroup}

%
% Sections and Subsections
% counters
% \count11=section
\countdef\sectno=11
\sectno = 0
% \count12=subsection
\countdef\sbsectno=12
\sbsectno = 0
% count16=subsubsectno
\countdef\ssbsectno=16
\ssbsectno=0
\def\section#1{\advance \sectno by 1 \sbsectno = 0
\sbsectno=0\ssbsectno=0\vskip 24pt
{\goodbreak
\noindent {\number\sectno.\ {#1}}}
\nobreak\vskip 12pt }
% heading of a sub-section
\def\subsection#1{\advance \sbsectno by 1
\ssbsectno=0
{\ifnum\count12=0\nobreak \else \medbreak \fi
\vskip 12 pt
\noindent {\number\sectno.\number\sbsectno. {\it #1}}}
\nobreak\vskip 12pt }
\def\subsubsection#1{\advance \ssbsectno by 1
\medbreak\vskip 12pt{
\noindent{\number\sectno.\number\sbsectno.\number\ssbsectno.
\it #1.\ }}}
%
% Figures
%
% \count13 = figure
\countdef\figno=13
\figno = 0
% figures: full width, top
% 1st argument:height with units, ex: 7.5cm
% 2nd argument:caption
\def\fig#1#2{\advance\figno by 1 \tolerance=10000
\setbox0=\hbox{\rm Fig. 00\ }
{\topinsert
\vskip #1 \smallskip\hangindent\wd0%
\noindent Fig.~\the\figno .\ #2
\vskip 12pt\endinsert}}
%
% Heading for acknowledgments
%
%

%
% Heading for list of references
%
\def\references
     {\countdef\refno=14
     \refno = 0\vskip 24pt
     {\noindent REFERENCES\par\nobreak}
      \parindent=0pt
      \vskip 12pt\frenchspacing}
\def\ref#1#2#3
   {\advance\refno by 1
    \item{\rm\number\refno. }\rm #1\ {\sl #2}
    \rm #3\par}
%
% fraction, alternative to \over
%
\def\frac#1#2{{#1 \over #2}}
%
% Renaming the dot under macro
%

%
% \d now used for differential d in mathematics
%
\def\d{{\rm d}}
%
% \e gives roman e for exponential e in mathematics
%

%
% \i gives roman i for square root of minus one in maths mode
% and dotless i in text mode
%
\def\i{\ifmmode{\rm i}\else\char"10\fi}
%
% et al
%

%
$\,\,\,$
\baselineskip=18pt
\bigskip
\centerline{\bf RENORMALIZATION GROUP APPROACH TO INTERACTING}
\centerline{{\bf CRUMPLED SURFACES: the hierarchical
recursion}$^{\ast}$}
\bigskip\bigskip\bigskip
\centerline{\it M.Cassandro$^{(\ast\ast)}$ and P.K.Mitter$^{(+)}$}
\bigskip\bigskip
\centerline{$^{(\ast\ast)}$Dipartimento di Fisica, I Universit\`a di Roma ``La
Sapienza''}
\centerline{P.le A.Moro 2, I-00185 Roma, (Italy)}
\bigskip\bigskip
\centerline{$^{(+)}$Laboratoire de Physique Th\'eorique et Hautes
Energies$^{1,2}$}
\centerline{Universit\'e Pierre et Marie Curie (Paris 6)}
\bigskip\bigskip\bigskip\bigskip\bigskip\bigskip\bigskip
\noindent
PAR-LPTHE 9346

\noindent{\underbar{Abstract}}
\bigskip
We study the scaling limit of a model of a tethered crumpled D-dimensional
random surface
interacting through an exclusion condition with a fixed impurity in
d-dimensional Euclidean
space by the methods of Wilson's renormalization group.

In this paper we consider a hierarchical version of the model and we prove
rigorously the
existence of the scaling limit and convergence to a non-Gaussian fixed point
for $1 \leq
D< 2$ and $\epsilon >0$ sufficiently small, where $\epsilon = D - (2-D) {d\over
2}$.

\footnote{}{$^{\ast}$~~~ Work supported
by the CNRS and in part by EC Science grant
SC$\ast$10394}
\vskip 1truecm

\footnote{}{$1$~ URA 280 au CNRS\par
2~  Postal address:  LPTHE,
Universit\'e Paris 6, Bo\^{\i}te 126, 4 Place Jussieu, 75252\par
Paris Cedex 05, France\par
e-mail: mitter@lpthe.jussieu.fr ~~~~cassandro@roma1.infn.it}

\bigskip \bigskip

\vfill\eject
\section{\underbar {Introduction}}
There has been much interest recently in the statistical mechanics of tethered
surfaces
and the associated crumpling transition. [See the contributions in [1] and
[2-7]]. The
underlying Hamiltonian has stretching and bending terms as well as an Edwards
interaction (generalized to surfaces) modelling self-avoidance [See the
contributions of
D.R.Nelson and others in [1]). In $[1-7]$ perturbative renormalisability is
assumed and the
fixed point in the crumpled phase governing long distance asymptotics is
calculated in
$\epsilon$-expansion in lowest order. To make further progress in the study of
renormalizability Duplantier, in [5], proposed the study of a simpler model
where the
Edwards interaction corresponding to self-avoidance is replaced by an
interaction with a
\underbar{fixed} impurity via an exclusion condition. The model is that of the
quantum field
theory of a fluctuating D-dimensional crumpled surface embedded in
d-dimensional Euclidean
space corresponding to an Euclidean classical action:

$$S = {1\over 2} \int d^D x \vert \nabla\vec\phi (x)\vert^2 + \tilde{g}_0 \int
d^D x
\delta^{(d)} (\vec\phi(x))\eqno(1.1)$$
which we can restrict to a finite volume. Here $\vec\phi(x) \varepsilon \Re^d$,
and
represents a point of a D-dimensional surface embedded in $\Re^d$. The Gaussian
term
corresponds to a so called phantom surface whose Haussdorf dimension is
identified with
$d_H = {2D\over 2-D} [1]$. The $\delta$-function interaction in (1.1) forbids
this surface
from touching a point (in this case the origin) of $R^d$. This is highly
singular,
and thus renormalizability is at stake for the field theory.

However F.David et al [7] have completed the perturbative renormalisation
programme to all
orders and, for $1 \leq D <2$, found the RG non-Gaussian fixed point in
$\epsilon$-expansion. Here $\epsilon = D -(2-D) {d\over 2}$. The purpose of
this article is
to show how the methods of Wilson's renormalization group [8], $(RG)$, can be
profitably
and rigorously applied to the study of the scaling limit of the Duplantier
model. The idea,
as usual, is to introduce an ultraviolet (UV) cutoff version of (1.1), and then
to remove
this cutoff (thus achieving the scaling limit) by thinning out degrees of
freedom. through
successive RG interations. However, in the present paper, we make one
simplification: we
replace the cutoff free field by its hierarchical version [Gallavotti
[9],[10]]. As is well
known the UV cutoff hierarchical free field retains the standard scaling
properties and
long-distance behaviour but eliminates non-localities in the RG and the
resulting RG
transformation is similar to the approximate recursion relation discussed by
Wilson [8].
To discuss the "more realistic" theory (without hierarchical approx) it is
necessary to take
care of the non localities that will arise. Standard techniques to handle this
problem are
the cluster and Mayer expansions, (see for instance, [10], [11], [12]). This
will be the
subject of a subsequent paper. But the RG analysis and convergence to a
non-Gaussian fixed
point is best seen first in the hierarchical framework where the underlying
mechanisms are
more transparent.

In this paper, starting from an UV cutoff version of (1.1) and in the
hierarchical
scheme, we will prove the existence of the scaling limit (UV
cutoff$\to\infty$), for $1
\leq D < 2$ and $\epsilon >0$ sufficiently small. In Section II the cutoff
version of the
model is presented and in Section III the RG and its \underbar{hierarchical
version}. We
have gone into some pedagogical detail in Section III for the uninitiated
reader.
In sections IV and V we give a rigorous proof of the convergence of the RG
iterations to a
non-Gaussian fixed point. We estimate the effective potential at every step
through
convergent expansions. We prove that the sequence of effective potentials
converges to a
fixed point. The strategy is similar to that of Gawedzki and Kupiainen [13,14]
in a different
context. The precise statement of our results is given by Theorem 1 at the
beginning of
section V. Some technical matters are left to the Appendices.

\vfill\eject
\section{\underbar{The UV cutoff model}}

We introduce a momentum space cutoff-function $F(p^2)$ where $F > 0, C^\infty$,
monotonic
decreasing, $F(p^2) \to 0$ as $p^2 \to\infty$ rapidly and $F(0)=1$. An example
of such
a cutoff function is
$$F(p^2)=e^{-p^2}$$
{\underbar {We assume}}:$\int d^D p F(p^2) <\infty$. Our cutoff free field
propagator
in momentum space is:
$$C_\Lambda (p) ={F(p^2/\Lambda^2)\over p^2}$$
where $\Lambda$ is the UV cutoff. We will choose $\Lambda=L^N$ (in fixed
units), so
$\Lambda\to \infty$ as $N\to \infty$.{\underbar {Here $L\geq 2$ is an
integer}}. The
corresponding cutoff action is:
$$S_\Lambda (\varphi) = {1\over 2} \int d^Dx\,\,\, (\nabla \varphi(x),
F({-\Delta\over
\Lambda^2})^{-1} \nabla\varphi(x))+ \tilde{g} _0(\Lambda) \int d^Dx
\,\,\delta^{(d)}_\Lambda
(\varphi(x))\eqno(2.1)$$
where $\varphi(x) \epsilon \Re^d$ and ($\cdot$ , $\cdot$) is the inner product
in $\Re^d$.
Here the cutoff dependence of $\tilde g_0(\Lambda)$ is to be chosen, and
$\delta_\Lambda^{(d)}$ is an approximating sequence such that
$\delta_\Lambda^{(d)} \to
\delta^{(d)}$, as $\Lambda\to\infty$, in the sense of distributions. We will
choose:

$$\delta_\Lambda^{(d)} (\varphi) = \Bigl({\tilde{\lambda}_0 (\Lambda)\over
2\pi}\Bigr)^{d\over 2} \,\,
e^{-\,{\tilde\lambda_0(\Lambda)\vert\varphi\vert^2\over
2}}\eqno(2.2)$$
where $\vert\cdot\vert$ is the Euclidean norm in $\Re^d$, $\tilde\lambda_0 > 0$
with the
property  $\tilde\lambda_0(\Lambda) \to \infty$ as $\Lambda\to\infty$. The
cutoff
dependence will be fixed presently.

{}From the free field piece of (2.1) we see that the canonical (engineering)
dimension of
$\varphi$ in mass units, is:

$$[\varphi] = {D-2\over 2}\eqno(2.3)$$
which means, from (2.2),

$$[\tilde\lambda_0] = 2-D\eqno(2.4)$$
and hence from (2.2) and (2.1) we have:

$$[\tilde{g}_0] = D - (2-D) {d\over 2} \equiv \varepsilon\eqno(2.5)$$
We can now introduce \underbar{dimensionless couplings} $\lambda_0, g_0$ via:

$$\eqalign{&\tilde{\lambda}_0(\Lambda) = \Lambda^{2-D}\,\,\lambda_0\cr
&\tilde{g}_0(\Lambda) = \Lambda^\varepsilon\,\,g_0\cr}\Bigr\}\eqno(2.6)$$
where $\lambda_0, g_0$ are \underbar{cutoff independent} and \underbar{held
positive}.
We shall hold $1\leq D < 2$ and $\varepsilon > 0$, so that
$\tilde\lambda_0(\Lambda)$ and $\tilde g_0(\Lambda) \to \infty$ as
$\Lambda\to\infty$.
(The marginal case $\varepsilon = 0$ corresponds to critical internal dimension
$D_{crit} =
{2d\over d+2}$ or critical external dimension $d_{crit} = {2D\over 2-D}$).

It is convenient to pass to``dimensionless'' fields $\Phi$ (unit cutoff) given
by:

$$\varphi(x) = \Lambda^{D-2\over 2} \Phi (\Lambda x)\eqno(2.7)$$
Substituting this in (2.1) and making use of (2.6), we get the unit cutoff
action:

$$\eqalign{S_1(\Phi) &= {1\over 2} \int d^D x\,\,\, (\nabla\Phi(x) F(-\Delta)
\nabla\Phi(x))\cr &+ \int d^D x \,\, v_0 (\Phi(x))\cr}\eqno(2.8)$$
where:

$$v_0(\Phi) = g_0 ({\lambda_0\over 2\pi}\Bigr)^{d\over 2}\,\,
e^{-\,{\lambda_0\over
2} \vert\Phi\vert^2}\eqno(2.9)$$

To reach the continuum limit ($\Lambda\to\infty$) starting from $S_\Lambda$ in
(2.1), is
equivalent to starting from the unit cutoff action (2.8), performing RG
iteractions $ln
\Lambda = N$ times and taking $N\to\infty$, [8]. As we shall see, \underbar{no
unstable
directions are encountered}, so that no further renormalization is necessary.
However
to speed up the convergence, and simplify the analysis, we will hold the
dimensionless
parameter $\lambda_0$ at a fixed value $\lambda_\ast$, which will be defined
later.

\vfill\eject
\section{\underbar{The R.G. and its hierarchical version}}

The partition function corresponding to the unit cutoff action (2.8) is given
by:
$$Z = \int d\mu _{C_1} (\Phi) e^{-V_o(\Phi)}\eqno(3.1)$$
where

$$\eqalign{&\mu_{C_1}(\Phi) = \matrix{d\cr \otimes\cr j=1\cr} \mu_{C_1}(\Phi_j)
\cr
&V_0(\Phi) = \int d^D x\,\,\,\,\,\,v_0(\Phi(x))\cr}\Bigr\}\eqno(3.2)$$
(see 2.9), and $\mu_{C_1}$ is the Gaussian measure of mean $0$ and unit cutoff
covariance
$C_1$, whose integral kernel is:

$$C_1(x-y) = \int {d^D p\over (2\pi)^D}\,\,e^{i\,p\cdot(x-y)}\,\,{F(p^2)\over
p^2}\eqno(3.3)$$
(3.1) is well defined, if we restrict ourselves to a finite volume
$\vartheta_N$, with
$\vert\vartheta_N\vert = L^{ND}\vert\vartheta\vert$. (If the original cutoff
action
(1.1) is held in fixed volume $\vartheta$, then passing to unit cutoff
increases
$\vartheta$ to $\vartheta_N$ as above).

\noindent
The sample field $\Phi$ is at least once differentiable, since $\int d^D p
F(p^2) <
\infty$.

To obtain a RG transformation we write

$$C_1 = C_{1/L} + \Gamma$$
where in momentum space,

$$\eqalign{&\tilde C_{1/L} (p) = {F(L^2 p^2)\over p^2}\cr
&\tilde\Gamma (p) = {F(p^2) - F(L^2 p^2)\over p^2}\cr}\Big\}\eqno(3.4)$$
and, correspondingly, write:

$$\Phi = \varphi + \zeta$$
as a sum of independent Gaussian random fields distributed with covariance
$C_{1/L},
\Gamma$ respectively. Correspondingly, the partition function Z can be written
as:

$$\eqalign{Z &= \int d\mu_{C_{1/L}} (\varphi) \int d\mu_\Gamma (\zeta)\,
e^{-V_0(\varphi +
\zeta)}\cr
&= \int d\mu_{C_1} (\Phi) \int d\mu_\Gamma (\zeta)\, e^{-V_0 (L^{{2-D\over
2}}\, \Phi
(L^{-1} \cdot) + \zeta)}\cr}$$
The RG transformation then is:

$$\eqalign{&V_0(\Phi) \to V_1 (\Phi),\cr
&e^{-V_1(\Phi)} = \int d\mu_\Gamma(\zeta) \, e^{-V_0 (L^{{2-D\over 2}}\, \Phi
(L^{-1}\cdot) + \zeta)}\cr}\eqno(3.5)$$
where $\mu_\Gamma$ is Gaussian measure with covariance $\Gamma$ (see 3.4). The
above
transformation is to be iterated $ln\Lambda = N$ times and $N\to\infty$, to
achieve the
continuum limit.

Note that $\Gamma$ has exponential decay in $x$-space, since $\tilde\Gamma (p)$
is
regular at $p=0$. Even if $V_0$ is local, $V_1$ will be not. But the RG
functional
integral, and thus $V_1$, can be studied by a high temperature expansion (for
this, for
other models, see for instance Gallavotti [10], [11],
D.C.Brydges [11], and the contributions of Brydges, Gallavotti, Gawedzki-
Kupiainen, Feldman et al in [12]).

In this paper we obviate this difficulty by introducing the hierarchical
approximation
to the cutoff free field, which enforces locality in the RG transformation and
keeps
scaling properties intact.

To this end we first introduce a sequence of independent Gaussian random
vectors
$\{\zeta_n\}^\infty_{n=0}$, $\zeta_n (x) \varepsilon \Re^d$ with covariance

$$\eqalign{&\langle \zeta^i_n (x) \zeta^j_m(y)\rangle =
\delta_{ij}\,\delta_{nm}\,
\Gamma_n (x-y)\cr
&\Gamma_n (x-y) = \Gamma \Bigl({x-y\over L^n}\Bigr)\cr}\eqno(3.6)$$
Then we observe that the unit cutoff free field $\Phi$, distributed according
to
covariance $C_1$, see (3.3), can be written as:

$$\Phi = \sum^\infty_{n=0}\, L^{n\cdot {(2-D)\over 2}}\, \zeta_n\eqno(3.7)$$
as can be checked by computing its covariance.

\noindent
Because of (3.6), the $\zeta_n$ are \underbar{almost} piecewise constant on
scale $L^n$,
in probability. To see this we use Tchebycheff's inequality:

$$\eqalign{P\{\vert\zeta_n(x) - \zeta_n(y) \vert >\delta\} &\leq {1\over
\delta}\,
E(\vert\zeta_n(x) - \zeta_n(y)\vert^2)\cr
&\leq\, {\vert x-y\vert^2\over \delta^2}\,\,E (\vert\nabla \zeta_n (\bar
x)\vert^2)\cr}$$
by the mean value theorem, since a sample field $\zeta_n$ is at least once
differentiable. Now,

$$\eqalign{E(\vert\nabla \zeta_n (\bar x)\vert^2) &= {d\over L^{2n}}\, \int
{d^D p\over
(2\pi)^D} (F(p^2) - F(L^2 p^2))\cr
&= {c\over L^{2n}}\,\,, {\rm by} (3.6).\cr}$$
Hence for $\vert x-y\vert \leq L^n$

$$P\{\vert \zeta_n(x) -\zeta_n(y)\vert > \delta\} \leq {c\over \delta^2}$$
which shows that $\zeta_n (x)$ is nearly constant on scale $L^n$ in
probability.

We also record:

$$\langle \vert \zeta_n (x)\vert^2 \rangle = d \gamma$$
where

$$\gamma = \int {d^D p\over (2\pi)^D} \tilde\Gamma (p) < \infty\eqno(3.8)$$
(we will evaluate this later for typical strongly cut off functions $F(p^2)$).

Observe also that (3.7), can be written as:

$$\Phi = \zeta_0 + L^{{(2-D)\over 2}} \bar\varphi$$
where

$$\bar\varphi = \sum^\infty_{n=0} L^{n\,{(2-D)\over 2}}
\,\zeta_{n+1}\eqno(3.9)$$
and it is easy to check, by computing the covariance, that

$$\bar\varphi (x) = \Phi ({x\over L})\eqno(3.10)$$

\bigskip
\noindent
\underbar{Hierarchical RG}

The hierarchical free field is modelled on (3.7) and the properties of
$\zeta_n$
explained before. Namely, we replace the Gaussian random vectors $\zeta_n$
which are almost
constant in probability on scale $L^n$ by Gaussian random vectors
$\zeta_{\Delta,n}$,
which are strictly constant on blocks $\Delta$ of size $L^n$. These random
vectors are
independent for distinct blocks of the same size, and also for blocks of
different size.
The independent Gaussian random vectors have the same covariance $\gamma$
(3.8).
Substituting these random vectors in (3.7), gives the hierarchical cutoff free
field.
Scaling properties are thus preserved.

More precisely, following Gallavotti [9], we introduce a sequence of compatible
pavings
$\{Q_n\}^\infty_{n=0}$, by blocks $\Delta \varepsilon Q_n$ of linear size $L^n$
of
$\Re^D$. Here $Q_n$ is a refinement of $Q_{n+1}$. To each block $\Delta
\varepsilon Q_n$
we associate an independent Gaussian random vector $\zeta_{n,\Delta}
\varepsilon \Re^d$
with covariance:

$$\langle \zeta^i_{n,\Delta}\,\,\zeta^j_{m,\Delta  '}\rangle = \delta_{ij}
\delta_{nm}
\delta_{\Delta\Delta '}\,\,\, \gamma\eqno(3.11)$$
Let $\Delta_n(x)$ be the unique block $\Delta_n \in Q_n$ with $x \in
\Delta_n$. The \underbar{hierarchical} cutoff free field is obtained by
replacing (3.7)
by:

$$\Phi (x) = \sum^\infty_{n=0}\, L^{n\,{(2-D)\over 2}}\,
\zeta_{\Delta_n(x)}\eqno(3.12)$$
For cutoff $\Lambda$, the fields $\varphi$ live in $\vartheta \subset \Re^D$
with
$\vert\vartheta\vert = L^{MD}$, for $\vartheta$ a hypercube of size $L^M$.
Hence as
observed after (3.3), the unit cutoff field $\Phi$ lives in hypercube of size
$L^{N+M}$.
Hence (3.12) should be strictly replaced by:

$$\Phi (x) = \sum^{N+M}_{n=0}\,\, L^{n\,{(2-D)\over 2}}\, \zeta_{\Delta_n
(x)}\eqno(3.12')$$
The $\zeta_{\Delta_n}$ are piecewise \underbar{constant}, \underbar{but all
scaling
properties are preserved}. We can write (3.12') as:

$$\Phi(x) = \zeta_{\Delta_0(x)} + L^{{2-D\over 2}}\, \bar\varphi_{\Delta_1}
(x)\eqno(3.13)$$
where,

$$\bar\varphi_{\Delta_1} (x) = \sum^{N+M-1}_{n=0} \, L^{n\,{(2-D)\over 2)}}\,
\zeta_{\Delta_{n+1} (x)}$$
$\Delta_1 \subset \Delta_2 \subset \Delta_3 \subset \ldots$. The subscript
$\Delta_1$ in
$\bar\varphi_{\Delta_1}$ emphasizes that $\bar\varphi$ is piecewise constant on
blocks
$\Delta_1 \varepsilon Q_1$. It is easy to see that $\bar\varphi_{\Delta_1} (x)
= \Phi
\Bigl({x\over L}\Bigr)$, since $x \varepsilon \Delta_{n+1} \Rightarrow {x\over
L}
\varepsilon \Delta_n$.

Our RG transformation (3.5) simplifies considerably for the hierarchical cutoff
field. (3.5)
reads:

$$e^{-V_1 (\Phi)} = \int \matrix{\Pi\cr {\Delta_0 \varepsilon Q_0}\cr}
d\mu_\gamma
(\zeta_{\Delta_0}) e^{-V_0 (\zeta + L^{{2-D\over 2}}\bar{\varphi}})$$
where

$$d\mu_\gamma (\zeta_{\Delta_0}) = (2\pi\gamma)^{-d/2}\,\, e^{-
{\vert\zeta_{\Delta_0}\vert^2\over 2\gamma}}\, d^d
\zeta_{\Delta_0}\,.\eqno(3.14)$$
Since $\Phi$ is piecewise constant over blocks $\Delta_0 \varepsilon Q_0$,

$$V_0 (\Phi) = \int d^D x \,\,\,\,v_0 (\Phi(x)) = \sum_{\Delta_0 \varepsilon
Q_0}\, v_0
(\Phi_{\Delta_0})\eqno(3.15)$$
where

$$\Phi_{\Delta_0} = \sum_{n\geq 0}\, L^{n\,{(2-D)\over 2}}\, \zeta_{\Delta_n},
\quad
\Delta_0 \subset \Delta_1 \subset \Delta_2 \subset \ldots$$
Hence, using (3.15) and (3.13),
$$\eqalign{e^{-v_0(\zeta + L^{{2-D\over 2}}\, \bar\varphi}) &= \matrix{\Pi\cr
{\Delta_0
\varepsilon Q_0}\cr}e^{-v_0(\zeta_{\Delta_0} + L^{{2-D\over 2}}\,
\bar\varphi_{\Delta_1}})\cr
& = \matrix{\Pi\cr {\Delta_1
\varepsilon Q_1}\cr}\matrix{\Pi\cr {\Delta_0
\subset \Delta_1}\cr{(\Delta_0 \varepsilon Q_0})\cr}e^{-v_0(\zeta_{\Delta_0} +
L^{{2-D\over
2}}\, \bar\varphi_{\Delta_1})}} $$
Plugging this into the RG transformation (3.14), we have
$$\eqalign{\bar{e}^{V_1 (\Phi)}  &= \matrix{\Pi\cr {\Delta_1
\varepsilon Q_1}\cr}
\matrix{\Pi\cr {\Delta_0
\subset \Delta_1}\cr{\Delta_0 \varepsilon Q_0}\cr}
\int d\mu_\gamma(\zeta_{\Delta_0})\,\,\,e^{-v_0(\zeta_{\Delta_0} + L^{{2-D\over
2}}\,
\bar\varphi_{\Delta_1}})\cr
& = \matrix{\Pi\cr {\Delta_1
\varepsilon Q_1}\cr} \Bigl[\int d\mu_\gamma(\zeta_{\Delta_0})
e^{-v_0(\zeta_{\Delta_0} + L^{{2-D\over 2}}\,
\bar\varphi_{\Delta_1})}\Bigr]^{L^D}\cr}\eqno(3.16) $$
since $\sharp\{\Delta_0$ bloks in $\Delta_1\} = L^D$.

\bigskip
\noindent
\underbar{We define the hierarchical RG transformation}:

$$e^{-v_1 (\bar\varphi_{\Delta_1})} = \Bigl[\int d\mu_\gamma
(\zeta_{\Delta_0}) e^{-v_0 (\zeta_{\Delta_0} + L^{{2-D\over 2}}
\bar\varphi_{\Delta_1})}\Bigr]^{L^D}\eqno(3.17)$$
Then from (3.16),

$$\eqalign{e^{-V_1(\Phi)} = \matrix{\Pi\cr \Delta_1 \varepsilon Q_1\cr}\,
e^{-v_1(\bar\varphi_{\Delta_1})} &= e^{- \sum_{\Delta_1 \varepsilon Q_1} v_1
(\bar\varphi_{\Delta_1})} = e^{-\sum_{\Delta_0 \varepsilon Q_0} v_1
(\Phi_{\Delta_0})}\cr
&= e^{- \int d^D x \,\,\,\,\,v_1 (\Phi(x))}\cr}\eqno(3.18)$$

\vfill\eject
\section{\underbar{Hierarchical RG iterations: the first step}}

We begin the study of the sequence of hierarchical RG iterations:

$$v_n (\varphi) \to v_{n+1} (\varphi), \quad \varphi \varepsilon \Re^d$$
where

$$v_{n+1} (\varphi) = - L^D\, ln \Bigl( (\mu_\gamma \ast e^{v_n}) (L^{{2-D\over
2}}
\varphi)\Bigr)\eqno(4.1)$$
where

$$\eqalign{(\mu_\gamma \ast e^{-v_n}) & (L^{{2-D\over 2}} \varphi) = \int
d\mu_\gamma
(\zeta) e^{-v_n (\zeta + L^{{2-D\over 2}} \varphi)}\cr
& d\mu_\gamma (\zeta) = (2\pi\gamma)^{-{d\over 2}}\, e^{-
{\vert\zeta\vert^2\over 2\gamma}}\, d^d \zeta\cr}\eqno(4.2)$$
with

$$v_0 (\varphi) = g_0 ({\lambda_0\over 2\pi})^{d\over 2}\, e^{-{\lambda_0\over
2}\,
\vert\varphi\vert^2}\eqno(4.3)$$
Note that $\gamma$, given by (3.8), can be evaluated to be: (use $F(p^2) =
e^{-p^2}$ as
cutoff function)

$$\gamma = {1\over 2^{D/2}} \cdot {2\over 2-D}\,\, (L^{2-D} - 1)$$
and so:

$$\gamma = O (L^{2-D}), \,\,\,{\rm for} \,\,D<2\,.\eqno(4.4)$$
{}From this, and 4.1 - 4.3, we immediately have analytic continuation in D for
$$1\leq D < 2$$.

To set the ball rolling, look at the first iteration $v_0 \to v_1$ in lowest
order in
$g_0$. We have

$$\eqalign{(\mu_\gamma \ast e^{-v_0})\, (L^{{2-D\over 2}} \varphi) &= 1 -
g_0\,\,\,\,
({\lambda_0\over 2\pi})^{{d\over 2}}\,\,\, (\mu_\gamma \ast
e^{-\lambda_0\vert\cdot\vert^2})\, (L^{{2-D\over 2}} \varphi) + O (g_0^2)\cr
&= 1 - g_0\,\,\, L^{-(2-D) {d\over 2}}\,\,\, ({\lambda_1\over 2\pi})^{{d\over
2}}\,
e^{-{1\over 2} \lambda_1 \vert\varphi\vert^2} + O (g^2_0)\cr}\eqno(4.5)$$
where:

$$\lambda_1 = {L^{2-D} \lambda_0\over 1+\gamma \lambda_0}\eqno(4.6)$$
and we have used the formula for Gaussian integration, for $u > 0$,

$$(\mu_\gamma \ast e^{- u\vert\cdot\vert^2})\, (L^{{2-D\over 2}} \varphi) =
{1\over
(1+\gamma u)^{d/2}}\, e^{-{1\over 2}\, {L^{2-D} u\over 1+\gamma u}
\vert\varphi\vert^2}\eqno(4.7)$$
Then:

$$\eqalign{v_1(\varphi) &= - L^D ln (\mu_\gamma \ast e^{-v_0} (L^{{2-D\over 2}}
\varphi))\cr
&=  L^\varepsilon\, g_0 \,\,\,({\lambda_1\over 2\pi})^{{d\over 2}}\,\,\,
e^{-{1\over 2} \lambda_1 \vert\varphi\vert^2} + O (g^2_0)\cr}\eqno(4.8)$$
where:

$$\varepsilon = D - (2-D) \cdot {d\over 2} > 0$$
Notice that we hold $D<2$, hence the map

$$\lambda_0 \to \lambda_1 = {L^{2-D} \lambda_0\over 1 + \gamma \lambda_0}$$
has a fixed point

$$\lambda_\ast = {L^{2-D} - 1\over \gamma}
={2^{-(2 - {D\over 2})}} (2-D) <1\eqno(4.9)$$
To simplify the further analysis,
and speed up the convergence, \underbar{we shall
choose} \underbar{the starting $\lambda_0 = \lambda_\ast$}, \underbar{and}

$$v_\ast (\varphi) = ({\lambda_\ast\over 2\pi})^{d\over 2}\,\,
e^{-{\lambda_\ast\over
2}\vert\varphi\vert^2}\eqno(4.10)$$
\underbar{and the starting interaction}:

$$v_0 (\varphi) = g_0\, v_\ast (\varphi)\eqno(4.11)$$

\bigskip
\noindent
\underbar{Iteration $v_0 \to v_1$}

\underbar{Notations}

$$\eqalign{&\beta = 2-D, \alpha = (2-D) {d\over 2} > 0, \quad d\geq 2\,.\cr
&\varepsilon = D - \alpha > 0, \quad 1 \leq D < 2\Bigr\}\cr}$$
We shall also write

$$\langle F\rangle_\varphi \equiv (\mu_\gamma \ast F)\, (L^{\beta/2}
\varphi)\qquad\qquad(\ast)$$
Then:

$$\eqalign{v_1 &= - L^D ln \Bigl( (\mu_\gamma \ast e^{-g_0 v_\ast})\,
(L^{\beta/2}
\varphi)\Bigr)\cr
&= - L^D ln \Bigl( 1 + \sum_{n\geq 1} {(-g_0)^n\over n!}\, \langle
v^n_\ast\rangle_\varphi\Bigr)\cr
&= - L^D \, \sum_{k\geq 1}\, {(-1)^{k-1}\over k}\, \Bigl(\sum_{n\geq 1}\,
{(-g_0)^n\over n!}\, \langle v^n_ \ast \rangle_\varphi \Bigr)^k\cr}$$
whence:

$$v_1 (\varphi) = g_0 L^D \sum_{k\geq 1}\, \sum_{\ell_1\ldots \ell_k}
(-g_0)^{\vert{\bf
\ell}\vert -1} c_k({\bf \ell}) \matrix{k\cr \Pi\cr j=1\cr} \langle v_\ast^
{\ell_j}\rangle_\varphi\eqno(4.12)$$
$$1 \leq l_j <\infty$$
where:

$$\eqalign{&{\bf \ell} = (\ell_1, \ldots, \ell_k),\,\,\, \vert{\bf \ell}\vert =
\sum^k_{j=1}
\ell_j\cr
&c_k ({\bf \ell}) = {(-1)^{k-1}\over k} \cdot {1\over \matrix{k\cr \Pi\cr
j=1\cr}
(\ell_j !)}\cr}\eqno(4.13)$$

\bigskip
\noindent
\underbar{Remark}:

By collecting terms of given power of $g_0$, we can write (4.12) as:

$$v_1 = L^D\, g_0 \sum_{n\geq 1}\, {(-g_0)^{n-1}\over n!}\, \langle
v^n_\ast\rangle_{conn}$$
and this is just the cumulant expansion.

By explicit Gaussian integration, using formula 4.7, together with the property
that
$\lambda_\ast = {1\over \gamma} (L^{2-D} - 1)$ is the fixed point of the
transformation
$\lambda\to\lambda ' = {L^\beta \lambda\over 1+\gamma\lambda}$, we obtain, for
large $L$,

$$\eqalign{\langle v^\ell_\ast\rangle_\varphi = L^{-\alpha} v_\ast
(\varphi)\,\,\, {1\over
\ell^{d/2}}\,\,\,\,\, &({\lambda_\ast\over 2\pi})^{(\ell -1){d\over
2}}\,\cdot\, (1+ (1 -
{1\over \ell}) L^{-\beta})\,\cdot\cr
&\cdot exp (- {\lambda_\ast\over 2} L^{-\beta} (1 - {1\over \ell})
\vert\varphi\vert^2)\cr}\eqno(4.14)$$
where strictly speaking $L^{-\beta} (1 - {1\over \ell})$ stands for
$O(L^{-\beta} (1 - {1\over
\ell}))$ and this will be understood.

\bigskip
\noindent
\underbar{Define}:

$$\eqalign{\tilde h_k &({\bf \ell}, g_0) = (-g_0)^{\vert{\bf \ell}\vert - 1}
c_k({\bf \ell}) d_k ({\bf \ell})\, ({\lambda_\ast\over 2\pi})^{(\vert{\bf
\ell}\vert -
1){d\over 2}}\cr
&d_k ({\bf \ell}) = \matrix{k\cr \Pi\cr j=1\cr} \,
\ell_j^{-d/2}\qquad\qquad\qquad\qquad\qquad\qquad\qquad\Biggr\}\cr
&f_k ({\bf \ell}) = \sum^k_{j=1}\, (1 -
{1\over \ell_j})\cr}\eqno(4.15)$$
with $c_k ({\bf \ell})$ given in (4.13).

Plugging in (4.14) into (4.12), we get:

$$v_1 (\varphi) = L^\varepsilon \, g_0 v_\ast (\varphi) G_1 (\varphi,
g_0)\eqno(4.16)$$
where

$$\eqalign{G_1 (\varphi, g_0) = \sum_{k\geq 1} &L^{-(k-1)\alpha}\,
\sum_{\matrix{\ell_1
\ldots \ell_k\cr 1\leq \ell_j < \infty\cr}} \tilde h_k ({\bf \ell}, g_0)\, (1 +
L^{-\beta} f_k ({\bf \ell})) \cdot\cr
&\cdot exp (- {\lambda_\ast\over 2} [(k-1) + L^{-\beta} f_k({\bf \ell})]
\vert\varphi\vert^2)\cr}\eqno(4.17)$$
This is a series of differentiable functions of the variable $\varphi^2 =
\vert\varphi\vert^2 \geq 0$. We have the uniform bound:

$$\vert G_1\vert \leq \sum_{k\geq 1} L^{-(k-1)\alpha}\, \sum_{\matrix{\ell_1
 \ldots \ell_k\cr 1 \leq \ell_j < \infty\cr}}\vert \tilde h_k ({\bf \ell}, g_0)
\vert
(1+L^{-\beta}k)\eqno(4.18)$$
and we claimn that the r.h.s. of (4.18) converges provided

$$\vert g_0\vert \leq ({\lambda_\ast\over 2\pi})^{-{d\over 2}}\, ln
(1+L^\alpha)\eqno(4.18')$$
Thus (4.17) converges uniformly. This shows that $G_1$ is a continuous function
of
$\varphi^2$, for sufficiently small $g_0$. Later we will see that it is
differentiable.

\noindent
\underbar{Proof of Claim}

\noindent
To check the uniform convergence, note that
$$\sum_{{l_1...L_k\atop 1 \leq l_j <\infty}} \vert \tilde{h}_k ({\underline l},
g_0)\vert
(1 + L^{-\beta} k)$$
$$\leq (1 + L^{-\beta}) \sum_{{l_1...lL_k\atop1 \leq l_j <\infty}} k \cdot
{(\vert
g_o\vert)^{(\sum^k_{j=1} l_{j-1})}\over k\cdot \Pi^k_{j=1} l_j!}
({\lambda_*\over
2\pi})^{(\sum^k_{j=1} lj_{-1}){d\over 2}}$$
$$= {(1+L^{-\beta})\over (\vert g_0\vert({\lambda_*\over 2\pi}^{d/2})}
(\sum_{l\geq 1}
{(\vert g_0\vert ({\lambda_*\over 2\pi})^{{d\over 2}})^l\over l!})^k$$
$$= {(1 +L^{-\beta})\over \vert g_0 \vert ({\lambda_*\over 2\pi})^{d/2}}
(e^{\vert g_0 \vert
({\lambda_*\over 2\pi})^{d/2}} -1)^k\eqno(*)$$
Putting in the bound (*) in (4.18) we see:
$$\vert G_1 \vert \leq (1+L^{-\beta}) [{e^{\vert g_0\vert ({\lambda_*\over
2\pi})^{d/2}}
-1\over \vert g_0 \vert ({\lambda_*\over 2\pi})^{d/2}}] \sum_{k\geq 1}
{(e^{\vert g_0 \vert
({\lambda_*\over 2\pi})^{d/2}} -1)^{k-1}\over L^{(k-1)\alpha}}$$
and the gemetric series on the $r\cdot h\cdot s$ converges provided:
$$\vert g_0 \vert \leq {1\over ({\lambda_*\over 2\pi})^{d/2}} \cdot ln (1+
L^\alpha)\eqno(4.18')$$
and the claim is proved.

\noindent
\underbar{Relevant and Irrelevant terms}

We now extract the relevant term which gives an effective coupling $g_1$, after
one $RG$
step, and a corresponding irrelevant term $I_1$, as follows.

\noindent
Define:

$$g_1 = L^\varepsilon g_0 G_1 (\varphi, g_0)\vert_{\varphi = 0}\eqno(4.19)$$
and

$$I_1 (\varphi, g_0) \doteq L^\varepsilon g_0 [G_1 (\varphi, g_0) - G_1 (0,
g_0)]\eqno(4.20)$$ so that we can write the effective potential $v_1$, 4.16,
as:

$$v_1 (\varphi) = v_\ast (\varphi) [g_1 +  I_1 (\varphi,
g_0)]\eqno(4.21)$$

(that $I_1$, is irrelevant will be seen presently)
{}From (4.17), and (4.19),
$$g_1 = L^\epsilon g_0 \sum_{n\geq 1} a_{n-1}\,\,\, g_0^{n-1}\eqno(4.22)$$
where

$$a_{n-1} = ({\lambda_\ast\over 2\pi})^{(n-1) {d\over 2}}\,(-1)^{n-1}
\sum^n_{k=1}\,
L^{-(k-1)\alpha} \, \sum_{\matrix{\ell_1 \ldots \ell_k\cr \vert {\bf \ell}\vert
= n\cr}} c_k
({\bf \ell}) d_k({\bf \ell}) (1+ L^{-\beta} f_k ({\bf \ell}))\eqno(4.23)$$
\underbar{and (4.22) converges absolutely for}: $\vert g_0\vert <
({\lambda_*\over
2\pi})^{-d/2} ln (1+ L^\alpha)$, as follows from (4.18, 4.18').

The first few coefficients are:

$$\eqalign{&a_0 = 1\cr
&a_1 = - {1\over 2}\,({\lambda_\ast\over 4\pi})^{{d\over 2}}\, (1 +
{L^{-\beta}\over 2} -
2^{d/2} L^{-\alpha}) \cdot\cr
&\,\,\,= - {1\over 2}\,({\lambda_\ast\over 4\pi})^{{d\over 2}}\, (1 + O
(L^{-\beta})) <
0\cr}\eqno(4.24)$$
{}From (4.17) we can write:
$$g_1 = L^\varepsilon g_0 (1+ a_1 g_0) + r (g_0)\eqno(4.25)$$
where
$$\vert r (g_0)\vert \leq c L^\varepsilon \vert g_0 \vert^3\eqno(4.26)$$
for $g_0$ sufficiently small.

Ignoring $r(g_0)$, we derive the approximate fixed point.:
$$\bar{g} = {L^\varepsilon -1\over L^\varepsilon (-a_1)} = 0(\varepsilon ln L)
>0\eqno(4.27)$$
Given $\varepsilon >0$, sufficiently small, choose the block size L very large
but bounded:
$$({2\over \varepsilon^{5/2}})^{{2\over \bar{\beta}}} < L < e^{{1\over
\varepsilon^{1/8}}}\eqno(4.28)$$
where (see later after 4.31)
$$\bar{\beta} = \beta - \varepsilon > 0$$
In particular (4.28) $\Rightarrow$
$$0 < \varepsilon < {1\over (log L)^8}$$
so that
$$\varepsilon \vert ln \varepsilon \vert < \bar{g} < C
\varepsilon^{7/8}\eqno(4.28a)$$
Chose the initial $g_o$ very close to $\bar{g}$:
$$\vert g_0 - \bar{g}\vert < {1\over 4}\,\,\,\, \varepsilon^{5/2}\eqno(4.29)$$
{}From (4.25) and
$$\bar{g} = L^\varepsilon \bar{g}\,\,\,\, (1 + a_1 \bar{g})$$
by taking the difference we can bound (use 4.29)
$$\vert g_1 - \bar{g}\vert \leq {1\over 2}~ \varepsilon^{5/2}\eqno(4.29a)$$
Hence, from (4.29) and (4.29a), we get
$$\vert \Delta g_0 \vert = \vert g_1 - g_0 \vert
\leq\varepsilon^{5/2}\eqno(4.30)$$
\underbar{We will show in Section V that (4.30) is not only stable under
iteration}

\noindent
\underbar{but contractive}:
$$\vert \Delta g_n \vert \leq k^{n-1}_* \vert \Delta g_0\vert$$
with $0 < k_* < 1$, and all subsequent effective couplings $g_n$ lie within an
$\varepsilon^{3/2}$ neighbourhood of $\bar{g}$.

We now turn to the {\underbar irrelevant term $I_1$ (4.20)}. $I_1$ vanishes at
$\varphi^2
=0$, and, by what we have shown for $G_1$, is continuous in $\varphi^2$. We
claim that it is
differentiable and satisfies the uniform bound:
$$\vert {dI_1\over d\varphi^2}\vert \leq {\bar{g}^2\over
L^{{\bar{\beta}/2}}}\eqno(4.31)$$
where $\bar{\beta} = \beta - \varepsilon > 0$ (for $\varepsilon = 0,
\bar{\beta}_c =
\beta_c = 2 -D_c = {4\over d+2} >0$. Hence for $\varepsilon >0$ sufficiently
small
$\bar{\beta} >0$ by continuity)

To see this take the derivative of (4.17) term by term and upper bound. Note
that the $k =
l=1$ term in 4.17 gives 1 and so does not contribute to $I_1$. Also, after
taking the
derivative, the $\varphi^2$-dependent terms (exponentials with negative
exponents) can be
bounded by 1. Hence
$$\eqalign{&\vert {dI_1\over d\varphi^2}\vert \leq L^\varepsilon~ \vert g_0
\vert~
{\lambda_*\over 2} L^{-\beta} (1+L^{-\beta})\cdot\cr
&\cdot\{\sum_{l\geq 2}  \vert \tilde{h}_1
(l,g_0)\vert + (1+L^{-\beta}) \sum_{k\geq 2} L^{-(k-2)} \cdot \sum_{{l_1\ldots
l_k\atop l_j
\geq 1}} k^2 \vert \tilde{h}_k (l,g_0)\vert \}\cr}$$
If we now plug in the expression (4.15, 4.13) for $\tilde{h}_k$, we can easily
verify that
the series in braces $\{\}$ converge for $\vert g_0 \vert < ({\lambda_*\over
2\pi})^{-d/2} ln
(1+L^\alpha)$ and $\{\}$ is $0(g_0)$.
We restrict our selves to ${1\over 4}\varepsilon^{5/2}$ neigbourhood of
$\bar{g}$ as in
(4.29). Then we have:
$$\vert{dI_1\over d\varphi^2}\vert \leq {C\bar{g}^2\over L^{\bar{\beta}}}$$
We use up a factor ${1\over L^{\bar{\beta}/2}}$ to bound the constant by 1. We
then get
(4.31).
Of course we can get a much stronger bound (as far as the field dependence is
concerned),
but we will not need it. In fact we shall replace (4.31) by a \underbar{weaker
bound}:
$$\vert {dI_1\over d\varphi^2}\vert \leq {\bar{g}^2\over L^{\bar{\beta}/2}}
e^{{\lambda_*\over 2} L^{-\beta} \vert \varphi\vert^2}\eqno(4.32)$$
and by integrating this from $0$ to $\varphi^2$, with $I_1 (0) =0$ we get:
$$\vert I_1 \vert \leq {\bar{g}^2\over L^{\bar{\beta}/2}} \varphi^2
\,\,\,\,e^{{\lambda_*\over 2} L^{-\beta} \vert \varphi\vert^2}\eqno(4.33)$$
The growth in $\varphi^2$ is harmless since, from (4.21), $I_1$ is always
multiplied by
$$v_* = ({\lambda_*\over 2 \pi}) ^{d/2} e^{-{\lambda_*\over 2} \vert
\varphi\vert^2}$$
and
$$\sup_{\varphi} \vert v_* I_1\vert \leq {\bar{g}^2\over L^{\bar{\beta}/2}}$$
\underbar{We will see in Section V that the bound (4.32) is stable
under iteration}.

We summarize what we have obtained after one iteration in the following
Proposition

\underbar{Proposition 1}

Let $\varepsilon >0$ be sufficiently small (4.28) and $\bar{g}$ be defined by
(4.27). Hold
$g_0$ so that
$$\vert g_0 - \bar{g}\vert \leq {1\over 4} \varepsilon^{5/2}$$
Then after 1 RG iteration
$$v_0 = g_0 v_* \to v_* (g_1 + I_1)$$
where $I_1 (0) =0$. $I_1 (\varphi)$ is $C^1$ in $\varphi^2$ and the following
bounds hold:
$$\vert \Delta g_0 \vert = \vert g_1 - g_0 \vert \leq \varepsilon^{5/2}$$
$$\vert {dI_1\over d\varphi^2}\vert \leq {\bar{g}^2\over L^{\bar{\beta}/2}}
e^{{\lambda_*\over 2} L^{-\beta} \vert \varphi\vert^2}$$
where $\bar{\beta} = \beta - \varepsilon > 0$.

\vfill\eject
\section{\underbar{Higher iterations and Convergence to non Gaussian fixed
point}}.

Let us write, in analogy to what we have obtained after one iteration, the
effective
potential $V_n$ after the $n th\,\, RG$ iteration in the form:
$$v_n (\varphi) = v_* (\varphi) [g_n + I_n (\varphi)]\eqno(5.1)$$
where $I_n (0) =0$, \underbar{$I_n$ being the irrelevant term}.
Recall, $v_* = ({\lambda_*\over 2 \pi})^{d/2} e^{-{\lambda_*\over 2} \vert
\varphi\vert^2}$
and define the \underbar{uniform norm} of $v_n$:
$$\Vert v_n \Vert = ({\lambda_*\over 2\pi})^{d/2} (\vert g_n \vert + \Vert I_n
\Vert_{v_*})\eqno(5.2)$$
where
$$\Vert I_n \Vert_{v_*} = \sup_{\varphi} \vert e^{{-\lambda_*\over 2} \vert
\varphi\vert^2} I_n (\varphi)\vert$$
In this section we will prove our main Theorem:

\vskip 10pt
\noindent
\underbar{Theorem 1} (Convergence to non-Gaussian fixed point)

As $n \to \infty , v_n \to v_\infty$ in the uniform norms $\Vert \cdot \Vert$
where
$$\Vert v_\infty \Vert = ({\lambda_*\over 2\pi})^{d/2} (\vert g_\infty \vert +
\Vert
I_\infty \Vert_{v_*})$$
and
$$\eqalign{&\vert g_\infty - \bar{g}\vert \leq \varepsilon^{3/2}\cr
&\Vert I_\infty \Vert_{v_*} \leq {4\bar{g}^2\over
L^{\bar{\beta}/2}}\cr}\eqno(5.3)$$
provided $\varepsilon > 0$ is sufficiently small and the block size L
sufficiently large
(the precise condition is (4.28) of Section IV)
Here,
$$\bar{g} = {L^\varepsilon - 1\over L^\varepsilon (-a_1)} = 0 (\varepsilon ln
L) >0$$
is the approximate fixed point of the first iteration, and
$$\bar{\beta} = \beta - \varepsilon = (2 - D) - \varepsilon >0\eqno(5.4)$$
(infact for $\varepsilon =0,\,\,\, \bar{\beta}_c = 2 - D_c = 2 - {2d\over d+2}
= {4\over d+2}
>0$ and hence, by continuity it follows that $\bar{\beta} >0$ for $\varepsilon
>0$ very
small).

\noindent
\underbar{Note that $g_\infty >0$ because of (5.3) and (4.28a)}.

\noindent
\underbar{Theorem 1} \underbar{thus states convergence of the sequence of
effective
potentials to a non-\-} $~$ \underbar{Gaussian fixed point close to the
approximate fixed
point of the first} iteration.

In order to prove Theorem 1 we shall bound the difference of successive
iterations:
$$\Delta v_n = v_{n+1} - v_n = v_* (\Delta g_n + \Delta I_n)\eqno(5.5)$$
(our strategy is similar to that of Gawedzki and Kupiainen in [13,14], in a
different
context)

To this end we shall make an inductive hypothesis (verified for $n=1$) for the
first n-steps
of RG iteration. As in section IV, (4.28) choose the block size L very large
but bounded
and $\varepsilon >0$ very small such that:
$${1\over ({1\over 2} L^{\bar{\beta}/2})^{2/5}} < \varepsilon < {1\over (ln
L)^8}\eqno(5.6)$$
This is easy to fulfill, as the reader can check. Since $\bar{g} = 0
(\varepsilon ln L)$,
the righthand inequality assures us that,
$$\bar{g}^2 < C \varepsilon^{7/4} = \varepsilon^{3/2} (C
\varepsilon^{1/4})\eqno(5.6a)$$
which is very small.
\underbar{Define}:
$$k_* = 1 - \varepsilon ln L + 10 \varepsilon^{3/2}\eqno(5.7)$$
Note that,
$$0 < k_* <1$$
Define also:
$$\delta (\varepsilon) = \varepsilon^{5/2}$$
$$\delta_* (\varepsilon) = \varepsilon^{3/2}\eqno(5.9)$$
and hold the initial coupling $g_0$ as in Section IV:
$$\vert g_0 - \bar{g} \vert < {1\over 4} \delta (\varepsilon)$$
Define:
$$\Delta g_l = g_{l+1} - g_l$$
$$\Delta I_l = I_{l+1} - I_l\eqno(5.10)$$

* \underbar{\it Inductive hypothesis}: For the first n-steps of the RG
iteration, the
effective potential $v_1 , v_2 ,\ldots,v_n$ satisfy the following,
\underbar{Property} $H_n$:

\noindent
For $l = 0,1,2,\ldots,n-1$
$$\eqalign{&{\rm (i)}~ \vert \Delta g_l \vert \leq k_*^l \delta
(\varepsilon)\cr
&{\rm (ii)}~ \vert {\d\over d\varphi^2} \Delta I_l \vert \leq c^{(l)}
e^{{\lambda_*\over 2}
L^{-\beta} \vert \varphi \vert^2}\cr}\eqno(5.11)$$
where:
$$c^{(0)} = {\bar{g}^2\over L^{\bar{\beta}/2}}$$
and for $l \geq 1$:
$$c^{(l)} = {\bar{g}\over L^{\bar{\beta}/2}} k_*^{l-1} \delta(\varepsilon) +
{\bar{g}^2\over
L^{\bar{\beta}/2}} ({1\over L^{\bar{\beta}/2}})^l\eqno(5.11a)$$

Note that for $n =1$, the inductive hypothesis is satisfied (Proposition 1 of
Section IV).

Also remark that Property $H_n \Rightarrow$ \underbar{the additional properties
$H_n^1$}:

\noindent
\underbar{Property} $H^1_n$
$${\rm (i a)}~~~ \vert g_n - \bar{g} \vert \leq \varepsilon^{3/2} = \delta_*
(\varepsilon)$$
$${\rm (ii a)}~~~ \vert {d\over d\varphi^2} I_n \vert \leq {4\bar{g}^2\over
L^{\bar{\beta}/2}} e^{{\lambda_*\over 2} \,^{L^{-\beta}} \vert \varphi
\vert^2}$$
$${\rm (ii b)}~~~ \vert \Delta I_{n-1}\vert \leq c^{(n-1)} \vert \varphi
\vert^2
e^{{\lambda_*\over 2} \,^{L^{-\beta}} \vert \varphi \vert^2}$$
$${\rm (ii c)}~~~ \vert I_n \vert \leq {4\bar{g}^2\over L^{\bar{\beta}/2}}
\vert \varphi
\vert^2 e^{{\lambda_*\over 2} \,^{L^{-\beta}} \vert \varphi \vert^2}$$
\underbar{Proof}. Start from $H_n$

\noindent
(i) $\Rightarrow$ (ia), since
$$\eqalign{&\vert g_n - \bar{g} \vert \leq \sum^{n-1}_{l=0} \vert \Delta g_l
\vert + \vert
g_0 - \bar{g}\vert \leq (\sum^{n-1}_{l=0} k^l_*) \delta(\varepsilon) + {1\over
2} \delta
(\varepsilon)\cr
&\leq {1\over 1-k_*} \varepsilon^{5/2} + {1\over 2} \varepsilon^{5/2}\cr
&\leq \varepsilon^{3/2}\cr}$$
Next, (ii) $\Rightarrow$ (ii a), since
$$\vert {d\over d\varphi} I_n \vert \leq \sum^{n-1}_{l=0} \vert {d\over
d\varphi^2} \Delta
I_l \vert \leq (\sum^{n-1}_{l-0} c^{(l)}) e^{{\lambda_*\over 2} L^{-\beta}
\vert
\varphi \vert^2}$$
and
$$\sum^{n-1}_{l=0} c^{(l)} \leq {\bar{g} \varepsilon^{5/2}\over
L^{\bar{\beta}/2} (1-K_*)} +
{\bar{g}^2\over L^{\bar{\beta}/2}} ({1\over 1- {1\over L^{\bar{\beta}/2}}})$$
$$\leq {4\bar{g}^2\over L^{\bar{\beta}/2}}$$

Moreover (ii a) $\Rightarrow$ (ii c), (integrate (ii a) with boundary condition
$I_n (0)
=0$).

Finally (ii) $\rightarrow$ (ii b) by the same reasoning.

\noindent
The main job of this section is to prove the following:

\noindent
\underbar{Theorem 2}. Suppose $v_1 , v_2,\ldots, v_n$ satisfies Property $H_n$
then
under RG iteration $v_{n+1}$ satisfies $H_{n+1}$.

Note that theorem 2 immediately implies Theorem 1:

\noindent
\underbar{Proof of Theorem 1} (given Theorem 2).

Since $v_1$ satisfies $H_1$ (Section IV, Proposition 1), Theorem 2$\Rightarrow
v_n$ has the
property $H_n$ for all $n$. In particular $H_n^1$ holds, all $n$.
Since, $0 < k_* < 1$, (i) of $H_n \Rightarrow \{g_n\}$ is a sequence whose
increments are
absolutely summable, and hence Cauchy and (ia) says that every $g_n$ lies
within an
$\varepsilon ^{3/2}$ ball of center $\bar{g}$.
Hence $g_n \to g_\infty$ and
$$\vert g_\infty - \bar{g} \vert \leq \varepsilon^{3/2}$$
The decrease of $v_*$ in $\varphi^2$ beats the growth allowed in (ii b) and (ii
c). From
(iic), the $I_n$ are uniformly bounded in the $\Vert \cdot \Vert_{v_*}$ norm.
$c^{(n-1)}$,
see (ii), goes to zero as $n \to \infty$, and is summable. Now from (ii b) it
follows that
the $I_n \to I_\infty$ in the $\Vert \cdot \Vert_{v_*}$ norm, and
$$\Vert I_\infty \Vert_{v_*} \leq {4\bar{g}^2\over L^{\bar{\beta}/2}}$$
{\it So Theorem 1 has been proved (given \underbar{Theorem 2})}

\hskip 10pt
The starting point for the proof of Theorem 2 will be the following formulae
which give
$\Delta v_n$ through the increments $\Delta g_n , \Delta I_n$.

To this end define:
$$A_n(\varphi) \dot{=} \sum_{l\geq 1} {(-1)^l\over l!} {<v_*^l (g_n +
I_n)^l>_\varphi\over L^{-\alpha} v_* (\varphi)}\eqno(5.13)$$
$$\Delta B_n(\varphi) \dot{=} \sum_{l\geq 1} {(-1)^{l-1}\over l!} \sum_{\vert
\leq m
\leq l} \pmatrix{l\cr m\cr} {<v_*^l (g_{n-1} + I_{n-1})^{l-m} (\Delta g_{n-1} +
\Delta
I_{n-1})^m\over L^{-\alpha} v_* (\varphi)}\eqno(5.14)$$
Then we have:
\underbar{Increment formulae}
$$\Delta g_n = L^\varepsilon (1 + L^{-\alpha} v_* (0) A_n (0))^{-1} \Delta B_n
(0)$$
$$\eqalign{{d\over d\varphi^2} &\Delta I_n = L^\varepsilon {d\over d\varphi^2}
[(1+
L^{-\alpha} v_* A_n)^{-1} \Delta B_n]\cr
&\Delta I_n (0) =0\cr}\eqno(5.15)$$
\underbar{\it Note that the above make sense provided the series (5.13 - 5.14)
converge and
$A_n$} $~$ \underbar{\it is sufficiently small. This will be seen to be true
presently
because of the inductive} $~$ \underbar{\it hypothesis and $H_n$}.

\noindent
\underbar{\it Proof of (5.15)}.
Start from
$$v_n = v_*~~~ (g_n + I_n)$$
and explicitly perfom the RG iteration $v_n \to v_{n+1}$.
We get (replace $v_0 = g_0 v_*$ in (4.12) of section IV by $v_n$)
$$v_{n+1}(\varphi) = L^D \sum_{k\geq 1} \sum_{l_1 \ldots l_k} (-1)^{\vert
{\underline l}
\vert -1} c_k ({\underline l}) \prod^k_{j=1} < v_*^{l_j} (g_n +
I_n)^{l_j}>_\varphi\eqno(*1)$$
and
$$< F>_\varphi = (\mu_\gamma * F) (L^{\beta/2} \varphi)$$
Now write:
$$g_n + I_n = (g_{n-1} + I_{n-1}) + (\Delta g_{n-1} + \Delta I_{n-1})$$
and expand binomially:
$$(g_n + I_n)^{l_j} = \sum^{l_j}_{m_j =0} \pmatrix{l_j\cr m_j\cr} (g_{n-1} +
I_{n-1})^{l_j
-m_j} (\Delta g_{n-1} + \Delta I_{n-1})^{m_j}$$
If we insert this in ($\ast$ 1) above for each j-factor, then the contribution
corresponding
to $m_j = 0, j =1,\ldots,k$ gives us back $v_n$. Hence,
$$\Delta v_n = L^D \sum_{k\geq 1} \sum_{{l_1 \ldots l_k\atop l_j \geq 1}}
(-1)^{\vert
{\underline l}\vert -1} c_k ({\underline l})\cdot$$
$$\cdot \sum_{{m_1\ldots m_k\atop 0\leq n_j \leq l_j}}' \prod^k_{j=1}
\pmatrix{l_j\cr
m_j\cr} <v_*^{l_j} (g_{n-1} + I_{n-1})^{l_j - m_j} (\Delta g_{n-1} + \Delta
I_{n-1})^{m_j}>\eqno(*2)$$
where $\sum'$ means at least one $m_j \geq 1$. We can write $\sum'$ as
$$\sum' = k \sum_{{m_1\ldots m_{k-1}\atop 0\leq m_j \leq l_j}} \sum_{1\leq m_k
\leq
l_k}$$
by symmetry.
Note that
$$k c_k ({\underline l}) = {(-1)^{k-1}\over \displaystyle{\prod^k_{j=1}}
l_j!}$$
from (4.13) of section 4.
With these replacements in ($\ast 2$), divide each j-factor there by
$L^{-\alpha} v_*$, and
compensate by multiplying within the k-sum by
$$(L^{-\alpha} v_*)^k = L^{-\alpha}v_* (L^{-\alpha} v_*)^{k-1}$$
The $L^{-\alpha} v_*$ can be factored out of (2) altogether, and $L^{D-\alpha}
v_* =
L^\varepsilon v_*$.

\noindent
Performing the $l_k$ sum gives $\Delta B_n$. Each $l_j$ sum, for $j =1,\ldots,
k-1$, gives
the identical contribution $A_n$ (the binomial series can be summed up again).
We thus get
$$\Delta v_n = L^\varepsilon v_* (\sum_{k\geq 1} L^{-\alpha (k-1)} v_*^{k-1}
(-1)^{k-1}
A_n^{k-1}) \Delta B_n$$
or
$$\Delta v_n = L^\varepsilon v_* (1+ L^{-\alpha} v_* A_n)^{-1} \Delta
B_n\eqno(*3)$$
$$= v_* [L^\varepsilon (1 + L^{-\alpha} v_* (0) A_n (0))^{-1} \Delta B_n (0)
+$$
$$+ \{L^\varepsilon (1 + L^{-\alpha} v_* A_n)^{-1} \Delta B_n - ({\rm
same}~{\rm at} \varphi
=0)\}]$$
$$= v_* [\Delta g_n + \Delta I_n]$$
\underbar{\it Formula (5.15) has been proved}.

We have to give bounds on various quantities appearing on the RHS of the
formulae given by
(5.15) expressing the increments $\Delta g_n , \Delta I_n$. To obtain these
bounds we shall
make repeated use of the following Proposition 2, \underbar{\it whose proof is
given in
the} $~$ \underbar{\it appendix A. This proposition gives a priori bounds on RG
integrals of
the type}

\noindent
\underbar{we encounter}.

\noindent
\underbar{\it Proposition 2}.

Let $1 \leq q \leq l$ and $F(\varphi)$ a $\mu_\gamma$ integrable $C^1$ function
of
$\varphi^2$, satisfying
$$F(0) =0$$
and
$$\vert {dF\over d\varphi^2}\vert \leq c_1 e^{{\lambda_*\over 2}~L^{-\beta}
\vert \varphi
\vert^2}\eqno(5.16)$$
where $c_1 > 0$ is a constant.

\noindent
Then
$$< v_*^l F^q >_\varphi = (\mu_\gamma * (v^l_* F)) (L^{\beta/2} \varphi)$$
is $C^1$ in $\varphi^2$, and there exists a constant $c_2 >0$, \underbar{\it
independent of
$L$}, such that:
$$~~~{\rm (i)}~~ \vert {d\over d\varphi^2} ({<v_*^l F^q>\over
L^{-\alpha}v_*})\vert \leq
c_1^q c_2 L^{-\beta} e^{{\lambda_*\over 2}~L^{-\beta} \vert \varphi
\vert^2}\eqno(5.17)$$
$${\rm (ii)}~~ \vert {<v_*^l F^q>\over L^{-\alpha} v_*}\vert_{\varphi =0} \vert
\leq c_1^q
c_2\eqno(5.18)$$
$$~~~{\rm (iii)}~~ \vert {<v_*^l F^q>\over L^{-\alpha} v_*}\vert \leq c_1^q c_2
[L^{-\beta}
\vert \varphi \vert^2 e^{{\lambda_*\over 2}~L^{-\beta} \vert \varphi \vert^2}
+1]\eqno(5.19)$$
\underbar{Remark}. There is a trivial generalization of this proposition where
we
consider two functions F and G with above properties with $d_1$ appearing in
(5.16),
instead of $c_1$, for G. Instead of $F^q$, take $F^{q_1} G^{q_2}, 1 \leq q_1 +
q_2 \leq l$.
Then (i) - (iii) continue to hold with the replacement $c_1^q \to c_1^{q_1}
d_1^{q_2}$,
and with the same constant $c_2$ independent of L.

\noindent
\underbar{Proof of Proposition 2}.

\underbar{This follows from Lemmas 1-6 in the appendix A}.

\underbar{\it Define now}:
$$F_{n,l,m,s} (\varphi) \dot{=} {<v_*^l (g_{n-1} + I_{n-1})^{l-m} (\Delta
I)_{n-1}^s>_\varphi\over L^{-\alpha} v_* (\varphi)}\eqno(5.20)$$
for $1 \leq m \leq l,~~ 0 \leq s \leq m$.

These objects will obviously be encountered in
bounding increments (5.15), (see (5.14)). Bounds on them are provided by the
following
Proposition 3, which follows immediately from Proposition 2 and Property $H_n$
of the
inductive hypothesis.

\noindent
\underbar{Proposition 3}:

Assume the inductive hypothesis with Property $H_n$.

Then there exists a constant $c_3 = 1 + 0 (\varepsilon^{1/2})$, such that
$${\rm (i)}~~ \vert {dF_{n,l,m,s}\over d\varphi^2}\vert \leq {c_2\over L^\beta}
(c^{(n-1)})^s
(c_3 \vert \bar{g}\vert)^{(l-m)} e^{{\lambda_*\over 2}~L^{-\beta} \vert \varphi
\vert^2}\eqno(5.21)$$
$${\rm (ii)}~~ \vert F_{n,l,m,s} \vert \leq c_2 (c^{(n-1)})^s (c_3 \vert
\bar{g}\vert)^{l-m}
[{\varphi^2\over L^\beta} e^{{\lambda_*\over 2} L^{-\beta} \vert
\varphi\vert^2}
+1]\eqno(5.22)$$
Here $c_2$, independent of $L$, is the constant of Proposition 2.

\noindent
\underbar{Proof}.
$$\left|{dF_{n,l,m,s}\over d\varphi^2} \right| \leq \sum^{l-m}_{p=0}
\pmatrix{l-m\cr p\cr}
\vert g_{n-1}\vert^p ~~~\left| {d\over d\varphi^2} {<v_*^l I_{n-1}^{(l-m) -p}
(\Delta
I_{n-1})^s\over L^{-\alpha} v_*}\right|\eqno(*)$$
{}From part (i) of Proposition 2, (5.17) together with the remark following it,
and part (ii)
of Property $H_n$ (5.11), (ii a) of $H'_n$ (5.12), the derivative term is
bounded by:
$${c_2\over L^\beta} (c^{(n-1)})^s ({4\bar{g}^2\over L^{\bar{\beta}/2}})^{(l-m)
-p}
{}~~e^{{\lambda_*\over 2}~L^{-\beta} \vert \varphi\vert^2}$$
So the $R \cdot H \cdot S$ of ($\ast$) is bounded by
$${c_2\over L^\beta} (c^{(n-1)})^s (\vert g_{n-1} \vert + {4\bar{g}^2\over
L^{\bar{\beta}/2}})^{(l-m)} ~~e^{{\lambda_*\over 2}~L^{-\beta} \vert \varphi
\vert^2}\eqno(**)$$
{}From part (ia) of $H'_n$ (5.12),
$$\vert g_{n-1} \vert \leq \vert \bar{g}\vert + \varepsilon^{3/2}$$
Hence,
$$\vert g_{n-1} \vert + {4\bar{g}^2\over L^{\bar{\beta}/2}} \leq c_3 \vert
\bar{g} \vert$$
where $c_3 = 1 + 0 (\varepsilon^{1/2})$, since $\bar{g}$ is $0(\varepsilon)$.

\noindent
Part (ii) of Proposition 3 now follows.

\noindent
Part (ii) follows in the same way, using part (iii) of Proposition 2, (5.19).

\noindent
\underbar{\it Proposition 3 has been proved}.

Using Propositions 2 and 3 and property $H_n$ of the inductive hypothesis, we
obtain easily
bounds on $A_n$ and $\Delta B_n$ and its derivatives (appearing in (5.15))
summarized in
the following proposition 4, whose proof is relegated to Appendix B.

\noindent
\underbar{Proposition 4}
$$\vert (1 + L^{-\alpha} v_* (0) A_n (0))^{-1} \vert \leq 1 + L^{-\alpha}
({\lambda\over
2\pi})^{d/2} \bar{g} + 3 \varepsilon^{3/2}\eqno(5.23)$$
$$\vert {dA_n\over d\varphi^2}\vert \leq {2c_2 c_3 \vert \bar{g}\vert\over
L^\beta}
e^{{\lambda_*\over 2} L^{-\beta} \vert \varphi \vert^2}\eqno(5.24)$$
$$\vert A_n \vert \leq 2 c_2 c_3 \,\,\vert \bar{g} \vert ({\varphi^2\over
L^\beta}
e^{{\lambda_*\over 2} L^{-\beta} \vert \varphi \vert^2} +1)\eqno(5.25)$$
$$\eqalign{&\vert \Delta B_n (0) \vert
\leq k_*^{n-1} \delta (\varepsilon) [1 - \bar{g}
({\lambda_*\over 4\pi})^{d/2} (1 + {1\over 2} L^{-\beta})\cr
&+ c'{\bar{g}\over L^{\bar{\beta}/2}} + 5 \varepsilon^{3/2}]\cr}\eqno(5.26)$$
$$L^\varepsilon \vert {d\over d\varphi^2} \Delta B_n \vert \leq {1\over 4}
[{\bar{g}\over L^{\bar{\beta}/2}} k_*^{n-1} \delta (\varepsilon) +
{\bar{g}^2\over
L^{\bar{\beta}/2}} ({1\over L^{\bar{\beta}/2}})^n ]\cdot e^{{\lambda_*\over
2}L^{-\beta}
\vert \varphi \vert^2}\eqno(5.27)$$
$$\eqalign{&\vert \Delta B_n \vert \leq c [\bar{g}
k_*^{n-1} \delta (\varepsilon) (1+{1\over k_*L^{\bar{\beta}/2}}) + \bar{g}^2
({1\over
L^{\bar{\beta}/2}})^n]\cdot\cr &\cdot ({\varphi^2\over L^\beta}
e^{{\lambda_*\over 2}
L^{-\beta} \vert \varphi \vert^2} +1)\cr}\eqno(5.28)$$
The bounds proven in Proposition 4 will now enable us to prove Theorem 2.

\noindent
\underbar{Proof of Theorem 2}

\noindent
\underbar{Claim 1}
$$\vert \Delta g_n \vert \cdot{\leq} k_*^n \delta (\varepsilon)$$
This is part (i) of property $H_{n+1}$ see (5.11).

\noindent
\underbar{Proof}: From (5.15), and the bounds (5.23, 5.26), we have
$$\vert \Delta g_n \vert \leq L^\varepsilon \vert (1 + L^{-\alpha} v_* (0) A_n
(0))^{-1}\vert \vert \Delta B_n (0)\vert$$
$$\leq k_*^{n-1} \delta (\varepsilon) L^\varepsilon [1- \bar{g}
({\lambda_*\over
4\pi})^{d/2} (1+ {1\over 2} L^{-\beta}) + {c'\bar{g}\over L^{\bar{\beta}/2}} +
5
\varepsilon^{3/2}]$$
$$\dot (1+ L^{-\alpha} ({\lambda_*\over 2\pi})^{d/2} \bar{g} + 3
\varepsilon^{3/2})$$
$$\leq k_*^{n-1} \delta (\varepsilon) \{L^\varepsilon [1 - \bar{g}
({\lambda_*\over
4\pi})^{d/2} (1+ {1\over 2} L^{-\beta} -2^{d/2} L^{-\alpha}) + 9
\varepsilon^{3/2}] +
{L^\varepsilon c' \bar{g}\over L^{\bar{\beta}/2}}\}$$
(we have used the fact, see 5.6a), that
$$\bar{g}^2 < (C \varepsilon^{1/4}) \varepsilon^{3/2}$$
Thus
$$\vert \Delta g_n\vert \leq k_*^{n-1} \delta (\varepsilon) \{ L^\varepsilon
[1- (-2a_1)
\bar{g} + 9 \varepsilon^{3/2}] +{L^\varepsilon c' \bar{g}\over
L^{\bar{\beta}/2}}\}$$
where $a_1$ is given by (4.25) of section IV.
{}From the definition of the approximate fixed point $\bar{g}$, see (4.27) of
Section IV:
$$\bar{g} = {L^\varepsilon -1\over L^\varepsilon (-a_1)}$$
we get
$$L^\varepsilon (1- (-2a_1) \bar{g}) =2- L^\varepsilon \leq 1 - \varepsilon ln
L\eqno(i)$$
Also, (see, (5.6))
$$L^\varepsilon \leq 1 + C \varepsilon^{7/8}$$
$$L^{-\beta /2} \leq \varepsilon^{5/2}$$
and
$$\bar{g} < C \varepsilon^{7/8}$$
so that
$$L^\varepsilon (9 \varepsilon^{3/2} + {c'\bar{g}\over L^{\bar{\beta}/2}}) < 10
\varepsilon^{3/2}\eqno(ii)$$
Using (i), (ii) in (*) above,
$$\vert \Delta g_n \vert \leq k_*^{n-1} \delta (\varepsilon) \{1 - \varepsilon
ln L +
10\varepsilon^{3/2}\}$$
$$\leq k_*^n \delta (\varepsilon)$$
by virtue of the definition (5.7).
\underbar{Claim 1 has been proved}.

\noindent
\underbar{Claim 2}
$$\vert {d\over d\varphi} \Delta I_n\vert \leq c^{(n)} e ^{{\lambda_*\over 2}
L^{-\beta}
\vert \varphi\vert^2}$$
$$c^{(n)} = {\bar{g}\over L^{\bar{\beta}/2}} k_*^{n-1} \delta (\varepsilon) +
\bar{g}^2
({1\over L^{\bar{\beta}/2}})^n$$
This is part (ii) of Property $H_{n+1}$ (see 5.11)

\noindent
\underbar{Proof}: Carrying out the derivative of the right hand side of the
equation giving
${d\over d\varphi^2} \Delta I_n$ in (5.15) gives:
$$\eqalign{\vert {d\over d\varphi^2} \Delta I_n \vert \leq &(1 - L^{-\alpha}
v_* \vert
A_n\vert )^{-1} \vert L^\varepsilon {d\over d\varphi^2} \Delta B_n\vert +\cr
&+\{(1 - L^{-\alpha} v_* \vert A_n \vert)^{-2} L^\varepsilon L^{-\alpha} (\vert
{dv_*\over
d\varphi^2} A_n \vert + v_* \vert {dA_n\over d\varphi^2}\vert)\vert \Delta
B_n\vert\cr}\}\eqno(**)$$
and $\alpha \geq \beta$, for $d\geq 2$

\noindent
{}From
$$v_* = ({\lambda_*\over 2\pi})^{d/2} e^{-{\lambda_*\over 2} \vert
\varphi\vert^2}$$
$${dv_*\over d\varphi^2} = - {\lambda_*\over 2} v_*$$
and the bounds (5.24, 5.25) on $A_n$ and its derivative, we easily derive
$$L^{-\alpha} v_* \vert {dA_n\over d\varphi^2}\vert \leq {2c_2 c_3 \vert
\bar{g}\vert\over L^{2\beta}} e^{-{\lambda_*\over 4}\vert
\varphi\vert^2}\eqno(iii)$$
$$L^{-\alpha} \vert {dv_*\over d\varphi^2} A_n \vert \leq {8 c_2 c_3 \vert
\bar{g}
\vert\over L^\beta} e^{-{\lambda_*\over 4}\vert \varphi \vert^2}\eqno(iv)$$
$$L^{-\alpha} v_* \vert A_n\vert \leq {4 c_2 c_3 \vert \bar{g} \vert\over
L^\beta}\eqno(v)$$
Now we bound the various terms in ($\ast \ast$) above.

\noindent
${\underline 1}$ Using (v) above and the bound (5.38) we get:
$$\vert (1 - L^{-\alpha} v_* \dot \vert A_n \vert)^{-1} \vert L^\varepsilon
{d\over
d\varphi^2} \Delta B_n \vert \vert$$
$$\leq (1 - {4 c_2 c_3 \vert \bar{g} \vert\over L^\beta})^{-1}~ {1\over 4}
[{\bar{g}\over
L^{\bar{\beta}/2}} k_*^{n-1} \delta (\varepsilon) + {\bar{g}^2\over
L^{\bar{\beta}/2}}
({1\over L^{\bar{\beta}/2}})^n]~ e^{{\lambda_*\over 2} L^{-\beta}}$$
$$\leq {1\over 2}~ [{\bar{g}\over L^{\bar{\beta}/2}} k_*^{n-1} \delta
(\varepsilon) +
{\bar{g}^2\over L^{\bar{\beta}/2}} ({1\over L^{\bar{\beta}/2}})^n]~
e^{{\lambda_*\over 2}
L^{-\beta} \vert \varphi \vert^2}$$
${\underline 2}$ Next using (iii) and (iv) above, togethere with the bound
(5.39) on $\Delta
B_n$ gives for the term in braces $\{\}$ in () above:
$$\vert \{\} \vert \leq (1 - {4c_2 c_3 \vert \bar{g} \vert\over
L^\beta})^{-1}\,\, ({1\over
L^{\bar{\beta}}})\,\, (10 c_2 c_3 \bar{g})\,\, e^{-{\lambda_*\over 4}\vert
\varphi
\vert^2}\cdot$$
$$\cdot 2c\,\, (k_*^{n-1} \delta (\varepsilon) \bar{g} + \bar{g}^2 ({1\over
L^{\bar{\beta}/2}})^n)\,\, ({\varphi^2\over L^\beta}\,\,\, e^{{\lambda_*\over
2}L^{-\beta}
\vert \varphi \vert^2} +1)\eqno(***)$$
We can trivially bound:
$$e^{-{\lambda_*\over 4} \vert \varphi^2 \vert}\,\,\,\, ({\varphi^2\over
L^\beta}~
e^{{\lambda_*\over 2}L^{-\beta} \vert \varphi \vert^2} +1)~ \leq c'\,\,\,
e^{{\lambda_*\over
2} L^{-\beta} \vert \varphi \vert^2}$$
We plug this into ($\ast\ast\ast$). Then from the overall factor $({1\over
L^{\bar{\beta}}})$
in ($\ast \ast \ast$) we use up ${1\over L^{\bar{\beta}/2}}$ to bound all
unnecessary
constants by $1/2$. We then get$$\vert \{\}\vert \leq {1\over 2} [{\bar{g}\over
L^{\bar{\beta}/2}} k_*^{n-1} \delta (\varepsilon) + {\bar{g}^2\over
L^{\bar{\beta}/2}}
({1\over L^{\bar{\beta}/2}})^n] e ^{{\lambda_*\over 2} L^{-\beta} \vert \varphi
\vert^2}$$
Putting together this bound, together with that in 1) above we get from
($\ast\ast$)
$$\vert {d\over d\varphi^2}\vert \Delta I_n \Vert \leq [{\bar{g}\over
L^{\bar{\beta}/2}} k_*^{n-1} \delta (\varepsilon) + {\bar{g}^2\over
L^{\bar{\beta}/2}}
({1\over L^{\bar{\beta}/2}})^n]e^{{\lambda_*\over 2}L^{-\beta} \vert \varphi
\vert}$$
$$= c^{(n)} e^{{\lambda_*\over 2} L^{-\beta} \vert \varphi \vert^2}$$
\underbar{Claim 2 has been proved}

\noindent
\underbar{This completes the proof of Theorem 2}.

As shown earlier Theorem 2 $\Rightarrow$ Theorem 1, and thus the scaling limit,
in the
hierarchical approximation, and convergence to a non-Gaussian fixed point has
been proved.

\centerline{\bf Acknowledgements}
We thank Dominique Mouhanna for bringing to our attention the work of David,
Duplantier and
Guitter. One of us (PKM) thanks David Brydges and Kristof Gawedzki for helpful
conversations on the Renormalization Group. In particular, he thanks Krystof
Gawedzki for
discussions related to the subject of this paper. He thanks G.Jona-Lasinio and
the
Dipartimento di Fisica, Universit\`a di Roma - La Sapienza, where part of this
work was
done, for hospitality.

The other (M.C.) thanks the CNRS for support which made his visit to Paris
possible and the
Laboratoire de Physique Th\'eorique et Hautes Energies, Universit\'e Paris VI,
for
hospitality.

\noindent
\underbar{Appendix A}

\noindent
This appendix is devoted to the proof of Proposition 2 of Section 5. First we
state a
useful representation.
Let $F(\varphi)$ be a $\mu_\gamma$-integrable $C^1$ function of $\varphi^2$.
$$< v_* F>_\varphi \dot{=} (\mu_\gamma * (v_* F)) (L^{\beta/2} \varphi)$$
Also, from Section IV and Gaussian integration
$$< v_*>_\varphi = L^{-\alpha} v_* (\varphi)$$
Thus,
$${<v_* F>_\varphi\over L^{-\alpha} v_* (\varphi)}= {\int {d^d \zeta\over (2
\pi
\gamma)^{d/2}} e^{-{1\over 2\gamma} \vert \zeta - L^{\beta/2} \varphi \vert^2}
v_* (\zeta)
F(\zeta)\over \int {d^d \zeta\over (2\pi \gamma)^{d/2}} e^{-{1\over 2\gamma}
\vert \zeta -
L^{\beta/2} \varphi \vert^2} v_* (\zeta)}$$
we can absorb,
$$v_* (\zeta) = ({\lambda_*\over 2 \pi})^{d/2} e^{-{\lambda_*\over 2} \vert
\zeta
\vert^2}$$
into the measure $d\mu_\gamma (\zeta)$ to obtain a new convolution kernel.

\noindent
Infact,
$$\eqalign{{d^d\zeta\over (2 \pi \gamma)^{d/2}} &e^{-{1\over 2\gamma} \vert
\zeta -
L^{\beta/2} \varphi \vert^2} v_* (\zeta)\cr
=({\lambda_*\over 2 \pi})^{d/2} \cdot {1\over (1+\gamma \lambda_*)^{d/2}} &{d^d
\zeta\over (2 \pi \sigma)^{d/2}} e^{-{1\over 2\sigma} (\zeta - L^{-\beta/2}
\varphi)^2}\cdot \cr}$$
$$\cdot e^{-{(1-L^{-\beta}) \vert \varphi \vert^2\over 2\sigma}}$$
where
$$\sigma = ({1\over \gamma} + \lambda_*)^{-1}$$
and we have used (Section IV), $1 + \gamma \lambda_* = L^\beta$.

\noindent
We therefore obtain:
$${<v_* F>_{(\varphi)}\over L^{-\alpha} v_* (\varphi)} = \int d\mu_\sigma
(\zeta -
L^{-\beta/2} \varphi) F(\zeta)\eqno(A1)$$
where $\mu_\sigma$ is the Gaussian measure of covariance
$$\sigma = ({1\over \gamma} + \lambda_*)^{-1} = {\gamma\over L^\beta} = 0 (1)$$
The representation (A1) will be used in the following.

\noindent
\underbar{Lemma 1}

\noindent
Let $F(\varphi)$ be a $\mu_\sigma$ integrable $C^1$ function of
$\varphi^2$.
Then,
$$\eqalign{&\vert {d\over d\varphi^2} ({<v_* F>_\varphi\over L^{-\alpha} v_*
(\varphi)})\vert\cr
\leq {1\over 2\sigma} e^{-{L^{-\beta} \vert \varphi \vert^2\over 2\sigma}}
L^{-\beta}
&\sum^\infty_{j=0} {1\over (2j)!} {(L^{-\beta} \vert \varphi \vert^2)^j\over
\sigma^{2j}}\cdot\cr
&\cdot \int d\mu_\sigma (\zeta) [(\zeta_1^2)^j + {(\zeta^2_1)^{j+1}\over (2
j+1)\sigma}]F(\zeta)\cr}\eqno(A2)$$
\underbar{Proof}: By invariance of F and the measure, the $L\cdot H\cdot
S\cdot$ of (A1) is
a function of $\varphi^2$. Apply
$${d\over d\varphi^2} = {1\over 2 \vert \varphi \vert^2} \vec{\varphi} \cdot
{\partial\over \partial \vec{\varphi}}$$
to the integral kernel in (A1). We then obtain:
$${d\over d\varphi^2} ({<v_*F>\over L^{-\alpha} v_*}) = {1\over 2\sigma}
[{1\over
\vert \varphi \vert^2} \int {d^d\zeta\over (2\pi \sigma)^{d/2}} e^{-{(\zeta -
L^{-\beta/2}\varphi)^2\over 2\sigma}} L^{-\beta/2} \vec{\varphi} \cdot
\vec{\zeta}\,\,
F(\zeta)$$
$$- L^{-\beta} \int {d^d\zeta\over (2\pi \sigma)^{d/2}} e^{-{(\zeta -
L^{-\beta/2}\varphi)^2\over 2\sigma}} F(\zeta)]\eqno(A3)$$
Since ${<v_* F>\over L^{-\alpha} v_*}$ is an inv. function of $\varphi^2$, we
can choose
coordinates:
$$\varphi = (\varphi_1 , 0,\ldots, o),~ \vert \varphi \vert^2 = \varphi^2$$
$$e^{- {(\zeta - L^{-\beta/2} \varphi)^2\over 2\sigma}} = e^{-{L^{-\beta} \vert
\varphi
\vert^2\over 2\sigma}} e^{-{\vert \zeta \vert^2\over 2\sigma}} e^{+L^{-\beta/2}
{\varphi_1
\zeta_1\over \sigma}}$$

$$= e^{{L^{-\beta}\vert \varphi \vert^2\over 2\sigma}} \sum^\infty_{j=0}
{1\over j!}
({L^{-\beta/2} \varphi_1\over \sigma})^j e^{-{\vert \zeta \vert^2\over
2\sigma}}
\zeta_1^j\eqno(*)$$
\underbar{Plugging ($\ast$) in to $(3)$ we get}:
$${d\over d\varphi^2} ({<v_* F>\over L^{-\alpha} v_*}) ={1\over 2\sigma}
e^{-{L^{-\beta}\vert \varphi \vert^2\over 2\sigma}} (J_1 - J_2)\eqno(A4)$$
\underbar{where}
$$J_1 = {1\over \vert \varphi \vert^2} \sum^\infty_{j=0} {1\over j!} {1\over
\sigma^j} (L^{-\beta/2}\varphi_1)^{j+1} \int d\mu_\sigma (\zeta) \zeta_1^{j+1}
F(\zeta)$$
Since $F$ is inv. function,and $\mu_\sigma$ is even in $\zeta_1$, only odd
$j\geq 1$
contribute. So with $j \to 2 j+1$,
$$J_1 = {1\over \vert \varphi \vert^2} \sum^\infty_{j=0} {1\over (j+1)!}
{1\over \sigma^{2j
+1}} (L^{-\beta/2} \varphi_1)^{2j +2} \int d\mu_\sigma (\zeta)\,\,\,
\zeta_1^{2j+1}
F(\zeta)$$
$$= L^{-\beta} \sum^\infty_{j=0} {1\over (2j)!} {(L^{-\beta} \vert \varphi
\vert^2)^j\over \sigma^{2j}} \int d\mu_\sigma (\zeta) {(\zeta_1^2)^{j+1}\over
\sigma(2j
+1)} F(\zeta)\eqno(A5)$$
\underbar{whereas},
$$J_2 = L^{-\beta} \sum^{\infty}_{j=0} {1\over j!} ({L^{-\beta/2}
\varphi_1\over \sigma})^j
\int d\mu_\sigma (\zeta)~ \zeta_1^j F(\zeta)$$
Only even $j$ contribute, so $j \to 2 j$
$$= L^{-\beta} \sum^\infty_{j=0} {1\over (2 j)!} {(L^{-\beta}\vert \varphi
\vert^2)^j\over \sigma^{2j}} \int d\mu_\sigma (\zeta) (\zeta_1^2)^j
F(\zeta)\eqno(A6)$$
Hence, from (A4), (A5), (A6) and we get:
$$\vert {d\over d\varphi^2} ({<v_* ( F>\over L^{-\alpha}v_*})\vert \leq {1\over
2\sigma}
e^{-{L^{-\beta}\vert \varphi \vert^2\over 2\sigma}} (vert J_1\vert + \vert
J_2\vert)$$
$$\leq {1\over 2\sigma} e^{-{L^{-\beta}\vert \varphi \vert^2\over 2\sigma}}
L^{-\beta}
\sum^\infty_{j=0} {1\over (2j)!} {(L^{-\beta} \vert \varphi \vert^2)^j\over
\sigma^{2j}}\cdot$$
$$\cdot \int d\mu_\sigma (\zeta) [(\zeta^2_1)^j + {(\zeta^2_1)^{j+1}\over (2j
+1)\sigma}] \vert F(\zeta)\vert\eqno(A7)$$
\underbar{Thus Lemma 1 has been proved}

\noindent
\underbar{Remark}: Lemma 1 will now be applied in the following pages to
special choices of
the function F.

\noindent
\underbar{Lemma 2}
\underbar{Assume}: $F(\varphi)$ is a $\mu_\gamma$ integrable $C^1$ function of
$\varphi^2 =
\vert \varphi \vert^2, F(0) =0$ and
$$\vert {dF\over d\varphi^2}\vert \leq c_1 e^{{\lambda_*\over 2} L^{-\beta}
\vert \varphi \vert^2}\eqno(A8*)$$
Then,
$$\vert {d\over d\varphi^2} ({<v_* F>\over L^{-\alpha} v_*}) \vert \leq c_1 c_2
L^{-\beta} e^{{\lambda_*\over 2} L^{-\beta} \vert \varphi \vert^2}\eqno(A8**)$$
\underbar{for some constant $c_2 > 1 , c_2$ is indept. of $L$}

\noindent
\underbar{Proof}:
We shall use Lemma 1.
Note that, from (A8*),
$$\vert F \vert \leq c_1 ({2\over \lambda_*} L^\beta) (e^{{\lambda_*\over 2}
L^{-\beta} \vert \varphi \vert^2}-1)\eqno(A9*)$$
Hence, from Lemma 1,
$$\vert {d\over d\varphi^2} ({<v_* F>\over L^{-\alpha} v_*})\vert \leq
{c_1\over 2\sigma}
L^{-\beta} \cdot ({2\over \lambda_*}) e^{-{L^{-\beta} \vert \varphi
\vert^2\over 2\sigma}} L^\beta \cdot$$
$$\cdot \eqalign{&\{\sum^\infty_{j=0} {1\over (2j)!} {(L^{-\beta} \vert \varphi
\vert^2)^j\over \sigma^{2j}} [(f (1,j) + {f(1,j+1)\over (2j+1)\sigma})\cr
&- (f(0,j)+ {f(0,j+1)\over (2j +1)\sigma})]\}\cr}\eqno(A10)$$
where:
$$f(\alpha, j) = \int d\mu_\sigma (\zeta) e^{{\lambda_*\over 2} \alpha
L^{-\beta} \vert
\zeta \vert^2} (\zeta_1^2)^j$$
$$= \int{d^d\zeta\over (2\pi \sigma)^{d/2}} e^{-{\sigma\over 2} (1- {\lambda_*
\alpha\over
\sigma} L^{-\beta}) \vert \zeta \vert^2} (\zeta_1^2)^j\eqno(A11)$$
$$= {1\over (1- {\lambda_* \alpha\over \sigma} L^{-\beta})^{d/2}} \cdot
{\sigma^j
\over (1 - {\lambda_* \alpha\over \sigma} L^{-\beta})^j} (2j -1)!!$$
$$(2j -1)!! = {(2j)!\over j ! 2^j} = (2j -1) (2j - 3)....1$$
Using, ${(2(j+1)-1)!!\over (2j +1)} = (2j -1)!!$
we obtain:
$$f(\alpha, j) + {f(\alpha, j +1)\over (2j +1)\sigma} = {1\over (1 - {\lambda_*
\alpha\over
\sigma} L^{-\beta})^{d/2}} {\sigma^j \over (1- {\lambda_*\alpha\over
\sigma}L^{-\beta})^j} \cdot {(2j)!\over j!2^j}\cdot$$
$$\cdot[1 + {1\over (1 - {\lambda_*\alpha\over \sigma} L^{-\beta})}]$$
$$[f(\alpha, j) + {f(\alpha, j+1)\over (2 j +1)\sigma}] - [...]_{\alpha =0}$$
$$= {(2j)!\over j! 2^j} \cdot \sigma^j [{1\over (1- {\lambda_*\alpha\over
\sigma}
L^{-\beta})^{d/2}} (1 + {1\over (1- {\lambda_* \alpha\over \sigma}
L^{-\beta})}) ({1\over
(1 + {\lambda_*\alpha\over \sigma} L^{-\beta})^j}) -2]\eqno(A12)$$
Hence the infinite sum in (A10)

$$\{\sum^\infty_{j=0} .....\} = e^{L^{-\beta} \vert \varphi \vert^2\over
2\sigma (1-
{\lambda_*\over \sigma} L^{-\beta})} \cdot {1\over (1 - {\lambda_* \over
\sigma}
L^{-\beta})^{d/2}} (1 + {1\over (1 - {\lambda_*\over \sigma} L^{-\beta})})$$
$$- 2~e^{L^{-\beta} \vert \varphi \vert^2\over 2\sigma}\eqno(A13)$$
We easily have
$${1\over (1-{\lambda_*\over \sigma}L^{-\beta})} \leq 1 + 2 {\lambda_*\over
\sigma} L^{-\beta}$$
$$1 + {1\over (1 - {\lambda_*\over \sigma} L^{-\beta})} \leq 2 (1 + {\lambda_*
\over \sigma} L^{-\beta})$$
and
$$1 - {1\over 1- {\lambda_*\over \sigma} L^{-\beta}} = - {\lambda_*\over
\sigma}
L^{-\beta} {1\over 1 - {\lambda_*\over \sigma} L^{-\beta}}$$
satisfies:
$$-2 {\lambda_*\over \sigma} L^{-\beta} \leq 1 - {1\over 1-{\lambda_*\over
\sigma} L^{-\beta}} \leq - {\lambda_*\over \sigma} L^{-\beta}$$
Hence from (A13)
$$\{\sum^\infty_{j=0} ....\} \leq 2\,\, e^{{L^{-\beta} \vert \varphi
\vert^2\over 2\sigma (1
- {\lambda_* \over \sigma} L^{-\beta})}} [ (1 + {4\lambda_*\over \sigma}
L^{-\beta}) -e^{{-\lambda_*\over \sigma^2} L^{-2 \beta} \vert\varphi
\vert^2}]$$
and, since
$$1 - e^{-{\lambda_*\over \sigma^2} L^{-2\beta} \vert \varphi \vert^2} \leq
{\lambda_*\over \sigma^2} L^{-2\beta} \vert \varphi \vert^2$$
$$\leq L^{-\beta} ({1\over \delta}) e^{\delta{\lambda_*\over \sigma^2}
L^{-\beta}
\vert \varphi \vert^2}$$
where $\delta$ has to be chosen.
$$\{\sum^\infty_{j=0} ....\} \leq 2 L^{-\beta} e^{{L^{-\beta} \vert \varphi
\vert^2\over
2\sigma (1- {\lambda_*\over \sigma} L^{-\beta})}}  [4{\lambda_*\over \sigma} +
{1\over \delta} e^{\delta {\lambda_*\over \sigma^2} l^{-\beta} \vert\varphi
\vert^2}]\eqno(A(14)$$
Plugging the bound (A14) in (A10) we get:
$$\vert {d\over d\varphi^2} ({<v_* F>\over L^{-\alpha} v_*})\vert$$
$$\leq {c_1\over 2 \sigma} L^{-\beta} {2\over \lambda_*} \cdot (2)
e^{{(\lambda_*)L^{-2\beta}\over \sigma^2} \vert \varphi\vert^2}[4
{\lambda_*\over \sigma}
+ {1\over \delta} e^{\delta {\lambda_*\over \sigma^2}L^{-\beta} \vert \varphi
\vert^2}]$$
$$\leq {2 c_1\over (\lambda_*) \sigma} L^{-\beta} [4 ({\lambda_*\over
\sigma}) e^{{\lambda_*\over 2} L^{-\beta} \vert \varphi \vert^2}+ {1\over
\delta}
e^{{\lambda_*\over 2} L^{-\beta} \vert \varphi \vert^2} (2 {\delta^2\over
\sigma^2}^2 +
{2L^{-\beta}\over \sigma^2})]$$
\underbar{Choose}: $\delta = {\sigma^2\over 2} (1 - {1\over
2}) = {\sigma^2\over 4} \cdot$ so that for L, suff. large \underbar{since
$\sigma = 0 (1)$}
$$({2\delta\over \sigma^2} + {2L^{-\beta}\over \sigma^2}) = 1- 1/2
+{2L^{-\beta}\over
\sigma^2} <1$$
and hence
$$\vert {d\over d\varphi^2} ({<v_*F>\over L^{-\alpha} v_*})\vert\leq {2c_1\over
(\lambda_*)\sigma} 4 ({\lambda_*\over \sigma} + {1\over \sigma^2})
e^{{\lambda_*\over 2}
L^{-\beta} \vert \varphi \vert^2} \cdot L^{-\beta}$$
\underbar{Choose} $c_2 = {8\over \sigma^2} (1+ {1\over \sigma\lambda_*})>1$

\noindent
Then:
$$\vert{d\over d\varphi^2} ({<v_* F>\over L^{-\alpha} v_*})\vert \leq c_1 c_2
L^{-\beta} e^{{\lambda_*\over 2}L^{-\beta} \vert \varphi \vert^2}$$
\underbar{and Lemma 2 has been proved}.

\noindent
\underbar{Lemma 3}

\noindent
For $1 \leq q \leq l,~ l\geq 2$
$$\vert {d\over d\varphi^2} {<v_*^l F^q>\over L^{-\alpha} v_*}\vert \leq c_1^q
L^{-\beta}
e^{{\lambda_*\over 2} L^{-\beta} \vert \varphi \vert^2}\eqno A(15*)$$
where F satisfies the hypothesis of Lemma 2

\underbar{Proof}

\noindent
Writing,
$${<v_*^l F^q>\over L^{-\alpha} v_*} = {<v_* (v_*^{l-1} F^q)>\over L^{-\alpha}
v_*}$$
apply Lemma 1, choosing for F, $(v_*^{l-1} F^q)$
Then
$$\vert {d\over d\varphi^2} {<v_*^l F^q>\over L^{-\alpha} v_*}\vert \leq
{1\over 2\sigma}
e^{-{L^{-\beta} \vert \varphi \vert^2\over 2\sigma}} L^{-\beta}\cdot$$
$$\cdot\sum^\infty_{j=0}{1\over (2j)!} {(L^{-\beta} \vert \varphi
\vert^2)^j\over \sigma^2
j}\int d\mu_\sigma (\zeta) [(\zeta_1^2)^j + {(\zeta^2_1)^{j+1}\over \sigma(2
j+1)}] \vert
(v_*^{l-1} F^q) (\zeta)\vert$$
$$v_*^{l-1} = ({\lambda_*\over 2 \pi})^{(l-1){d\over 2}}
e^{-{\lambda_*\over 2} (l-1) \vert \varphi \vert^2}\eqno A(16)$$
{}From (A9*), in the bound on $F(\zeta)$ following the hypothesis of Lemma 2,
$$\vert F (\varphi)\vert \leq c_1 \cdot ({2\over \lambda_*} L^\beta)
e^{{\lambda_*\over 2} L^{-\beta} \vert \varphi \vert^2} (1- e^{-{\lambda_*\over
2}
L^{-\beta} \vert \varphi \vert^2)}$$
$$\leq c_1 \cdot e^{{\lambda_* \over 2} L^{-\beta}\cdot \vert \varphi
\vert^2}$$
$$\leq c_1 \cdot {2\over \lambda_* \delta} \cdot
e^{{\lambda_*\over 2} (\delta + L^{-\beta}) \vert \varphi \vert^2}$$
\underbar{where $\delta$ is to chosen, $\delta >0$}

\noindent
Hence
$$\vert v_*^{l-1} (\zeta) F^q (\zeta)\vert \leq c_1^q [{2\over
\lambda_*\delta} ({\lambda_*\over 2 \pi})^{(l-1) {d\over 2}}]^q$$
$$e^{-{\lambda_*\over 2} [(l-1) - q (\delta + L^{-\beta})]\vert \zeta
\vert^2}\eqno A(17)$$
Choose:
$$\delta = {1\over 2} - L^{-\beta}\eqno A(18)$$
Then,
$$(l-1) - q (\delta + L^{-\beta}) \geq (l-1) - l (\delta + L^{-\beta}){\rm
since}~ q \leq l$$
$$= l [1 - {1\over l} - \delta - L^{-\beta}]$$
$$= l [{1\over 2} - {1\over l}]\geq 0,~ {\rm since}\, l\geq 2$$
Hence, $(l-1) - q (\delta + L^{-\beta}) \geq 0$
Using this in (A17), get:
$$\vert v_*^{l-1} (\zeta) F^q (\zeta) \vert \leq c_1^q [{4\over \lambda_* (1-2
L^{-\beta})} ({\lambda_*\over 2\pi})^{{d\over 2}}]^q\eqno A(19)$$
(have used $0 <\lambda_* \leq 1$, as is easy to show).

\noindent
Using the bound (A19) we have from (A16)
$$\vert {d\over d\varphi^2} {<v_*^l F^q>\over L^{-\alpha}v_*}\vert \leq {1\over
2\sigma}
e^{-{L^{-\beta} \vert \varphi \vert^2 \over 2\sigma}} L^{-\beta} c_1^q [{4
({\lambda_*\over 2\pi})^{d\over 2}\over \lambda_* (1-2 L^{-\beta})}]^q\cdot$$
$$\cdot\sum^\infty_{j=0} \{{1\over (2j)!} {(L^{-\beta} \vert \varphi
\vert^2)^j \over \sigma^{2j}} \int d\mu_\sigma (\zeta) [(\zeta^2_1)^j
{(\zeta^2_1)^{j+1}\over (2j +1)^\sigma}]\}\eqno A(20)$$
The integral is computed as in the proof of Lemma 2, and we get:
$$2 \cdot {(2j)!\over j 2^j} \sigma^{\dot{J}}$$
so that the sum
$$\sum^\infty_{j=0} \{\} = 2 e^{L^{-\beta} \vert \varphi \vert^2\over
2\sigma}$$
We then get from (A20)
$$\vert {d\over d\varphi ^2} {<v_*^l F^q>\over L^{-\alpha} v_*}\vert \leq c_1^q
L^{-\beta}
\{{1\over \sigma} [{4 ({\lambda_*\over 2\pi})^{d/2}\over \lambda_* (1 -2
L^{-\beta})}]^q\}$$
Look at the constant in $\{\}$. Since $0 <\lambda_* \leq 1 , d\geq 3$
$$\{\} \leq {1\over \sigma} [{4\over (2 \pi)^{3/2}} \cdot {1\over (1 - 2
L^{-\beta})}]^q\leq 1\eqno(*)$$.
$$1\leq q \leq l$$
since, $\sigma ={\gamma\over L^\beta} = 2^{\beta/2} {(1-L^{-\beta})\over
\beta}$, (and this
is regular as $\beta \to 0$).
Hence:
$$\vert {d\over d\varphi^2} {<v_*^l F^q>\over L^{-\alpha}v_*}\vert \leq c_1^q
L^{-\beta}$$
$$\leq c_1^q L^{-\beta} e^{{\lambda_*\over 2} L^{-\beta} \vert \varphi
\vert^2}\eqno
A(21)$$
\underbar{and Lemma 3 has been proved}.

\noindent
\underbar{Remark}.

\noindent
We can collect Lemma 2 and Lemma 3 into a single Lemma.

\noindent
\underbar{Lemma 4}

Let $1 \leq q \leq l$, and $F(\varphi)\,\,\, a\,\,\mu_\gamma$ integrable $C^1$
function in
$\varphi^2$, satisfying the bound
$$\vert {dF\over d\varphi^2}\vert \leq c_1 e^{{\lambda_*\over 2} \vert \varphi
\vert^2}\eqno(*)$$
Then $\exists c_2 >1$, \underbar{independent of L}, such that
$$\vert {d\over d\varphi^2} ({<v_*^l F^q>\over L^{-\alpha} v_*})\vert \leq
c_1^q c_2
L^{-\beta} e^{{\lambda_*\over 2} L^{-\beta} \vert \varphi \vert^2}\eqno
A(22)$$
\underbar{Proof} This follows from Lemmas 3 and 4.

\noindent
We need two further elementary bounds.

\noindent
\underbar{Lemma 5}

Let F satisfy the conditions of Lemma 2, which are the same in Lemmas 2-4,
Then, for $1
\leq q \leq l$, there exists a constant $c_2$ independent of $L$ such that
$$\vert {<v_*^l F^q>_\varphi\over L^{-\alpha} v_* (\varphi)}\vert_{\varphi =0}
\vert \leq
c_1^q c_2 \eqno A(23)$$
\underbar{Proof}
$$\vert {<v_*^l F^q>_\varphi \over L^{-\alpha} v_* (\varphi)}\vert_{\varphi =0}
\vert \leq
\int d\mu_\sigma (\zeta)~ v_*^{l-1} (\zeta)~ \vert F^q (\zeta)\vert$$
(i) $l =q =1$. Then, using (A9*),
$$\int d\mu_\sigma (\zeta) \vert F (\zeta)\vert \leq c_1 ({2\over
\lambda_*}L^\beta)
[(\int d\mu_\sigma (\zeta) e^{{\lambda_*\over 2} L^{-\beta} \vert \zeta
\vert^2})-1]$$
$$= c_1 ({2\over \lambda_*} L^\beta) [ (1 - {\lambda_*\over \sigma}
L^{-\beta})^{d/2} -1]$$
$$\leq c_1 c_2 $$
(ii) $1 \leq q \leq l, l\geq 2$. use (A19) and (*), between (A20) and (A21), to
get
$$\int d\mu_\sigma (\zeta) \vert v_*^{l-1} F^q (\zeta) \vert \leq c_1^q c_2$$
Putting (i) and (ii) together we get (A23). \underbar{Lemma 5 has been proved}.

Finally integrate (A22) from $0$ to $\varphi^2$ and use (A23) to obtain:

\noindent
\underbar{Lemma 6}
$$\vert {<v_*^l F^q>_\varphi\over L^{-\alpha} v_* (\varphi)}\vert \leq c_1^q
c_2
[L^{-\beta} \vert \varphi \vert^2 e^{{\lambda_*\over 2}L^{-\beta} \vert \varphi
\vert^2} +1]\eqno A(24)$$
Lemma 4,5,6 take care of Proposition 2.

\noindent
\underbar{Proposition 2 has been proved}. The remark after Proposition 2 is a
trivial
extension of the above.

\noindent
{\bf Appendix B}

\noindent
\underbar{Proof of Proposition 4}

\noindent
\underbar{1. Bounds on $A_n$ and derivatives}

Starting from (5.13) we can write:
$$A_n (0) = - {<V_* (g_n + I_n)>\over L^{-\alpha} v_*} \vert_{\varphi =0} +
R_1$$
$$ = - g_n - {<v_* I_n>\over L^{-\alpha} v_*}\vert_{\varphi =0} + R_1$$
where,
$$R_1 = \sum_{l\geq 2} {(-1)^l\over l!} {<v_*^l (g_n +I_n)^l >\over L^{-\alpha}
v_*}\vert_{\varphi =0}$$
Hence,
$$\vert A_n (0) + g_n \vert \leq \vert{<v_* I_n>\over
L^{-\alpha}v_*}\vert_{\varphi=0} \vert
+ \vert R_1 \vert$$
The first term on $R\cdot H\cdot S$ is bounded above by:
$$c_2 {4\bar{g}^2\over L^{\bar{\beta}/2}}$$
(use part (ii), 5.18, of Proposition 2 and (ii a) of property $H_n^1$ (5.12)).

\noindent
On the other hand,
$$\vert R_1 \vert \leq \sum_{l\geq 2} {1\over l!} \sum^l_{m=0} \pmatrix{l\cr
m\cr} \vert
g_n \vert^{l-m} \vert {<v_*^l I_n^m>\over L^{-\alpha}
v_*}\vert_{\varphi=0}\vert$$
$$\leq \sum_{l\geq 2} {1\over l!} \sum^l_{m=0} \pmatrix{l\cr m\cr} \vert g_n
\vert^{l-m}
c_2 ({4\bar{g}^2\over L^{\bar{\beta}/2}})^m$$
$$= c_2 \sum_{l\geq 2} {1\over l!} (\vert g_n \vert + {4\bar{g}^2\over
L^{\bar{\beta}/2}})^l$$
Now the serie converges. Use (ia) of proporty $H_n1$ (5.12) to deduce:
$$\vert R_1 \vert \leq 2 c_2 \bar{g}^2$$
Hence,
$$\vert A_n (0) + g_n \vert \leq 6 c_2 \bar{g}^2$$
so, using (ia) of $H'_n$, get
$$- \bar{g} - 2 \varepsilon^{3/2} \leq A_n (0) \leq - \bar{g} + 2
\varepsilon^{3/2}$$
Finally, from above and
$$(1 + L^{-\alpha} v_* A_n)^{-1} = 1 - L^{-\alpha} v_* A_n + {(L^{-\alpha} v_*
A_n)^2\over
1 + L^{-\alpha} v_* A_n}$$
deduce,
$$\vert (1 + L^{-\alpha} v_* (0) A_n (0))^{-1} \vert \leq 1 + L^{-\alpha}
({\lambda_*\over
2\pi})^{d/2} \bar{g} + 3 \varepsilon^{3/2}\eqno B(1)$$
(B1) is just (5.23).

\noindent
\underbar{Next we bound}: ${dAn\over d\varphi^2}$.
Starting from (5.13)
$$\vert {dAn\over d\varphi^2}\vert \leq \sum_{l\geq 1} {1\over l!} \sum^l_{m=0}
\pmatrix{l\cr m\cr} \vert g_n \vert^{l-m} \vert {d\over d\varphi^2} {<v_*^l
I_n^m>\over
L^{-\alpha} v_*}\vert$$
Use proposition 2 and (ii a), 5,12, of property $H_n'$, by the inductive
hypothesis. Then:
$$\vert {dAn\over d\varphi^2} \vert \leq \sum_{l\geq 1} {1\over l!}
\sum^l_{m=0}
\pmatrix{l\cr m\cr} \vert g_n \vert^{l-m} {c_2\over L^\beta} ({4\bar{g}^2\over
L^{\bar{\beta}/2}})^l e^{{\lambda_*\over 2} L^{-\beta} \vert \varphi
\vert^2}=$$
$${c_2\over L^\beta} \sum_{l\geq 1} {1\over l!} (\vert g_n \vert +
{4\bar{g}^2\over
L^{\bar{\beta}/2}})^l e^{{\lambda_*\over 2} L^{-\beta}}$$

\noindent
The series converges. Now use (ia) of $H_n'$, 5.12, to deduce
$$\vert {dAn\over d\varphi^2}\vert \leq {2c_2 c_3 \vert \bar{g}\vert\over
L^\beta}
e^{{\lambda_*\over 2}L^{-\beta} \varphi^2}\eqno B(2)$$
Finally, using part (iii) of Proposition 2, (5.19), and $H_n'$ we get in the
same way:
$$\vert A_n \vert \leq 2 c_2 c_3 \vert \bar{g}\vert ({\varphi^2\over L^\beta}
e^{{\lambda_*\over 2} L^{-\beta} \vert \varphi \vert^2} +1)\eqno B(3)$$
(B2) and (B3) are (5.24) and (5.25) respectively.

\noindent
{\underbar{2. Bounds on $\Delta B_n$ and derivatives}}

In (5.14) we make the binomial expansion:
$$(\Delta g_{n-1} +\Delta I_{n-1})^m = \sum^m_{s=0} \pmatrix{m\cr s\cr} (\Delta
g_{n-1})^{m-s} (\Delta I_{n-1})^s\eqno B(4)$$
to obtain:
$$\Delta B_n = \sum_{l\geq 1} {(-1)^{l-1}\over l!} \sum_{1 \leq m \leq l}
\pmatrix{l\cr
m\cr} \sum^m_{s=0} \pmatrix{m\cr s\cr} (\Delta g_{n-1})^{m-s} F_{n,l,m,s}$$
where $F_{n,l,m,s}$ is given by (5.20).

We separate out the contribution of the terms with $s=0$, (called $\Delta
C_n$), and the
contribution $s\geq 1$, (called $\Delta D_n$). $\Delta D_n$ has thus at least
one
irrelevant term $\Delta I_{n-1}$. We get
$$\Delta C_n = (\Delta g_{n-1}) \sum_{l\geq 1} {(-1)^{l-1}\over l!} \sum_{1
\leq m \leq
l} \pmatrix{l\cr m\cr} (\Delta g_{n-1})^{m-1} F_{n,l,m,0}$$
$$\eqalign{&= (\Delta g_{n-1}) [\sum_{l\geq 1} {(-1)^{l-1}\over (l-1)!}
F_{n,l,1,0} + \cr
&(\Delta g_{n-1})\sum_{l\geq 2}{(-1)^{l-1}\over l!} \sum_{2 \leq m \leq l}
\pmatrix{l\cr
m\cr} (\Delta g_{n-1})^{m-2} F_{n,l,m,0}]\cr}\eqno B(5)$$
and
$$\Delta D_n = \sum_{l\leq 1} {(-1)^{l-1}\over l!} \sum_{1 \leq m \leq l}
\pmatrix{l\cr
m\cr} \sum^m_{s=1} \pmatrix{m\cr s\cr} (\Delta g_{n-1})^{m-s} F_{n,l,m,s}\eqno
B(6)$$
and
$$\Delta B_n = \Delta C_n + \Delta D_n\eqno B(7)$$
First we give a bound on $\Delta C_n (0)$, which we write (see B 5),
$$\Delta C_n (0) = (\Delta g_{n-1}) [1 - F_{n,2,1,0} (0) +R_2 + R_3]$$
where
$$R_2 = \sum_{l\geq 3} {(-1)^{l-1}\over (l-1)!} F_{n,l,1,0} (0)$$
$$R_3 = \Delta g_{n-1} \sum_{l\geq 2} {(-1)^{l-1}\over l!} \sum_{2 \leq m \leq
l}
\pmatrix{l\cr m\cr} (\Delta g_{n-1})^{m-2} F_{n,l,m,0}(0)$$
Now,
$$F_{n,2,1,0}(0) = {<v_*^2 (g_{n-1} + I_{n-1})>\over L^{-\alpha}
v_*}\vert_{\varphi =0}$$
$$= g_{n-1} ({\lambda_*\over 4\pi})^{d/2} (1 + {1\over 2} L^{-\beta}) + {<v_*^2
I_{n-1}>\over L^{-\alpha} v_*}\vert_{\varphi =0}$$
{}From the inductive hypothesis and (i a) of $H_n'$ (5.12),
$$\vert g_{n-1} - \bar{g} \vert \leq \varepsilon^{3/2}$$
{}From (ii a) of $H'_n$ and part (i), 5.17, of Proposition 2,
$$\vert {< v_*^2 I_{n-1}>\over L^{-\alpha}v_*}\vert_{\varphi =0} \vert \leq c_2
{4\bar{g}^2\over L^{\bar{\beta}/2}}$$
Hence
$$\vert 1 - F_{m,2,1,0} (0) \vert \leq 1 - \bar{g} ({\lambda_*\over
4\pi})^{d/2} (1 +
{1\over 2} L^{-\beta}) + 2\varepsilon^{3/2}$$
We have from part (ii), (5.22), of Proposition 3,
$$\vert R_2 \vert \leq \sum_{l \geq 3} {1\over (l-1)!} \vert F_{n,l,1,0}
(0)\vert$$
$$\leq c_2 \sum_{l\geq 3} {1\over (l-1)!} (c_3 \vert \bar{g} \vert)^{l-1} \leq
2 c_2 (c_3
\bar{g})^2$$
$$\leq 2\varepsilon^{3/2}$$
since the series converges.

Next, always using Proposition 3, and part (i), (5.11), of property $H_n$ of
inductive
hypothesis:
$$\vert R_3 \vert \leq \vert \Delta g_{n-1} \vert \sum_{l\geq 2} {1\over l!}
\sum_{2\leq m
\leq l} \pmatrix{l\cr m\cr} \vert \Delta g_{n-1} \vert^{m-2} \vert
F_{n,l,n,0}(0) \vert$$
$$\leq k_*^{n-1} \delta(\epsilon) \sum_{l\geq 2} {1\over l!} \sum^l_{m=2}
\pmatrix{l\cr
m\cr} (k_*^{n-1} \delta (\varepsilon))^{m-2} (c_3 \vert \bar{g} \vert)^{l-m}$$
$$\leq 2 k_*^{n-1} \delta (\varepsilon)$$
$$\leq \varepsilon^{3/2}$$
since the series converges.
Using the above bounds we get:
$$\vert \Delta C_n (0)\vert \leq \vert \Delta g_{n-1} \vert [\vert 1 -
F_{n,2,1,0}^{(0)}
\vert + \vert R_2 \vert + \vert R_3 \vert]$$
or
$$\vert \Delta C_n (0) \vert \leq k_*^{n-1} \delta (\varepsilon) [1 - \bar{g}
({\lambda_*\over 4\pi})^{d/2} (1+{1\over 2} L^{-\beta}) + 5
\varepsilon^{3/2}]\eqno B(8)$$
Next we bound the derivative of $\Delta C_n$.

\noindent
Starting from (B5),
$$\eqalign{&\vert{d\over d\varphi^2} \Delta C_n \vert \leq \vert \Delta g_{n-1}
\vert [
\sum_{l\geq 2} {1\over (l-1)!} \vert {d\over d\varphi^2} F_{n,l,1,0}\vert +\cr
&+ \vert \Delta g_{n-1} \vert \sum_{l\geq 2} {1\over l!} \sum^l_{m=2}
\pmatrix{l\cr m\cr}
\vert \Delta g_{n-1} \vert^{m-2}\cr
&\cdot \vert {d\over d\varphi^2} F_{n,l,m,o} \vert]\cr}$$
Now use part (i), (5.21), of Proposition 3 and from inductive hypothesis part
(i) of $H_n$,
(5.11), to get:
$$\eqalign{&\vert {d\over d\varphi^2} \Delta C_n \vert \leq k^{n-1}_* \delta
(\varepsilon)
{c_2\over L^\beta} [\sum_{l\geq 2} {1\over (l-1)!} (c_3 \bar{g})^{l-1} +\cr
&+ k_*^{n-1} \delta(\varepsilon) \sum_{l\geq 2} {1\over l!} \sum^l_{m=2}
\pmatrix{l\cr
m\cr} (k_*^{n-1} \delta (\varepsilon))^{m-2}\cdot\cr
&\cdot (c_3 \bar{g}^{l-m}]\cdot e^{{\lambda_*\over 2}L^{-\beta} \vert \varphi
\vert^2}\cr
&\leq k_*^{n-1} \delta (\varepsilon) {c_2\over L^\beta} c_3 \bar{g}
[(1+0(\bar{g})) +
k_*^{n-1} \delta (\varepsilon) (1+0 (\bar{g}))]\cdot\cr
&\cdot e^{{\lambda_*\over 2} L^{-\beta} \vert \varphi\vert^2}\cr}$$
since the series converges. Hence,
$$\vert {d\over d\varphi}^2 \Delta C_n \vert \leq k_*^{n-1} \delta
(\varepsilon) {c_2 c_3
\bar{g}\over L^\beta} (1+ \varepsilon^{1/2})^2 e^{{\lambda_*\over 2} L^{-\beta}
\vert
\varphi \vert^2}\eqno B(9)$$
similarly using part (ii) of Proposition 3 and $H_n$,
$$\vert \Delta C_n \vert \leq k^{n-1}_* \delta (\varepsilon) c_2 c_3 \bar{g}
(1+
\varepsilon^{1/2})^2 [{\varphi^2\over L^\beta} e^{{\lambda_*\over 2} L^{-\beta}
\vert
\varphi \vert^2} +1]\eqno B(10)$$
Next we shall bound $\Delta D_n$ and its derivative.
Starting from (B6), and using part (i) of Proposition 3, (5.21) and part (i) of
$H_n$,
(5.11), we get:
$$\vert {d\over d\varphi^2} \Delta D_n \vert \leq {c_2\over L^\beta}
[\sum_{l\geq 1} {1\over
l!} \sum_{1 \leq n \leq l} \pmatrix{l\cr m\cr} \sum^m_{s=1} \pmatrix{m\cr s\cr}
(c^{(n-1)})^s (c_3 \bar{g})^{l-m}]\cdot e^{{\lambda_*\over 2} L^{-\beta} \vert
\varphi
\vert^2}$$
The series converges, and we deduce:
$$\vert {d\over d\varphi^2} \Delta D_n \vert \leq {c_2\over L^\beta} \cdot
c^{(n-1)} c
e^{{\lambda_*\over 2} L^{-\beta} \vert \varphi \vert^2}\eqno B(11)$$
In the same way, using part (ii) of Proposition 3 and part (i) of $H_n$ we get:
$$\vert \Delta D_n \vert \leq c_2 c^{(n-1)} c[{\varphi^2\over L^\beta}
e^{{\lambda_*\over
2} L^{-\beta} \vert \varphi \vert^2} + 1]\eqno B(12)$$
whence,
$$\vert \Delta D_n (0) \vert \leq c c_2 c^{(n-1)}\eqno B(13)$$
Now recall, (5.11 a) of $H_n$,
$$c^{(n-1)} = {\bar{g}\over L^{\bar{\beta}/2}} k_*^{n-2} \delta (\varepsilon) +
{\bar{g}^2\over L^{\bar{\beta}/2}} ({1\over L^{\bar{\beta}/2}})^{n-1}$$
Because of (5.6),
$${1\over L^{\bar{\beta}/2}} < {1\over 2} \epsilon^{5/2} = {1\over 2} \delta
(\varepsilon) <
k_* \delta (\varepsilon)$$
Hence,
$$c^{(n-1)} \leq k_*^{n-1} \delta (\varepsilon) [{\bar{g}\over k_*
L^{\bar{\beta}/2}} +
{\bar{g}^2\over L^{\bar{\beta}/2}}]$$
Hence, from (B 13),
$$\vert \Delta D_n (0)\vert ~ \leq k_*^{n-1} \delta (\varepsilon)
[{c'\bar{g}\over
L^{\bar{\beta}/2}} +\bar{g}^2]\eqno B(14)$$
{}From the above bounds on $\Delta C_n$ and $\Delta D_n$ we get bounds on
$$\Delta B_n = \Delta C_n + \Delta D_n$$
{}From (B8) and (B14) we get
$$\vert \Delta B_n (0) \vert \leq k_*^{n-1} \delta (\varepsilon) [1 - \bar{g}
({\lambda_*\over 4\pi})^{d/2} (1+ {1\over 2} L^{-\beta}) + {c' \bar{g}\over
L^{\bar{\beta}/2}} + 5\varepsilon^{3/2}]\eqno B(15)$$
This is (5.26).
On the otherhand, from (B9) and (B11), we have,
$$\vert L^\varepsilon {d\over d\varphi^2} \Delta B_n\vert \leq [{c_2c_3'
\bar{g}\over L^{\bar{\beta}}} k_*^{n-1} \delta (\varepsilon) + {c_2 c\over
L^{\bar{\beta}}}
c^{(n-1)}] e^{{\lambda_* \over 2} L^{-\beta} \vert \varphi \vert}$$
$$\leq {c_2 c'\over
L^{\bar{\beta}/2}} [{\bar{g}\over L^{\bar{\beta}/2}} k_*^{n-1} \delta
(\varepsilon) +
{1\over L^{\bar{\beta}/2}} c^{(n-1)}]e^{{\lambda_*\over 2} L^{-\beta} \vert
\varphi\vert}$$
Now,
$${1\over L^{\bar{\beta}/2}} c^{(n-1)} = {1\over L^{\bar{\beta}/2}}
({\bar{g}\over
L^{\bar{\beta}/2}} k_*^{n-2} \delta (\varepsilon) + {\bar{g}^2\over
L^{\bar{\beta}/2}}
({1\over L^{\bar{\beta}/2}})^{n-1})$$
$$= {\bar{g}\over L^{\bar{\beta}}} k_*^{n-2} \delta (\varepsilon) +
{\bar{g}^2\over
L^{\bar{\beta}/2}} ({1\over L^{\bar{\beta}/2}})^n$$
Plugging this into the previous inequality,
$$\vert L^\varepsilon {d\over d\varphi^2} \Delta B_n \vert \leq {c_2 c'\over
L^{\bar{\beta}/2}} [{\bar{g}\over L^{\bar{\beta}/2}} k_*^{n-1} \delta
(\varepsilon) (1+
{1\over k_* L^{\bar{\beta}/2}})+$$
$$+ {\bar{g}^2\over L^{\bar{\beta}/2}} ({1\over L^{\bar{\beta}/2}})^n] \cdot
e^{{\lambda_*\over 2} L^{-\beta} \vert \varphi \vert}$$

We can now use the overall factors $L^{-\bar{\beta}/2} $ to bound unecessary
constants by
$1/4$. We thus get:
$$\vert L^\varepsilon {d\over d\varphi^2} \Delta B_n \vert \leq {1\over 4}
[{\bar{g}\over
L^{\bar{\beta}/2}} k_*^{n-1} \delta (\varepsilon) + {\bar{g}^2\over
L^{\bar{\beta}/2}}
({1\over L^{\bar{\beta}/2}})^n]\cdot e^{{\lambda_*\over 2} L^{-\beta} \vert
\varphi
\vert^2}\eqno B(16)$$
which is (5.27).
Finally from (B10) and (B12), we obtain similarly:
$$\vert \Delta B_n \vert \leq c ~[\bar{g} k_*^{n-1} \delta (\varepsilon) (1+
{1\over k_*
L^{\bar{\beta}/2}})^n + \bar{g}^2 ({1\over L^{\bar{\beta}/2}})^n]\cdot
({\varphi^2\over
L^\beta} e^{{\lambda_*\over 2}L^{-\beta} \vert \varphi \vert^2} +1)\eqno
B(17)$$
which is (5.28).

\noindent
\underbar{This completes the proof of Proposition 4}.

\vfill\eject
\references
\ref{D.R.Nelson, T.Piron and S.Weinberg (eds):}{ Statistical Mechanics of
Membranes and
Surfaces, World Scientific, Singapore,}{ 1989.}

\ref{M.Kardar and D.R.Nelson:}{}{ Phys. Rev. Lett {\bf 58} (1987) 1298, Phys.
Rev. Lett
{\bf 58} (1987) 2280, Phys. Rev. A {\bf 38} (1988) 966}

\ref{J.A.Aronowitz and T.C.Lubensky,}{ Europhys.}{ Lett {\bf 4} (1987), 395}

\ref{B.Duplantier,}{}{ Phys. Rev. Lett {\bf 58} (1987) 2733}

\ref{B.Duplantier,}{}{ Phys. Rev. Lett {\bf 62} (1989) 2337}

\ref{B.Duplantier, T.Hwa and M.Kardar,}{}{ Phys. Rev. Lett {\bf 64}, (1990)
2022}.

\ref{F.David, B.Duplantier and E.Guitter:}{}{ Phys. Rev. Lett. {\bf 70} (1993)
2205, Nucl.
Phys. B 394 (1993), 555}

\ref{J.Kogut and K.G.Wilson:}{}{ Phys. Report {\bf 12} C (1974) 75}

\ref{G.Gallavotti}{ Memorie dell'Accademia dei Lincei}{ XV, (1978) 23}

\ref{G.Gallavotti}{ Rev. of Modern Phys}{ {\bf 57} (1985), 471}

\ref{G.Gallavotti}{ Lausanne Lectures}{ (3$^o$ cycl.de la Suisse Romande 1990)}
{ }
{D.C.Brydges:} { Lausanne Lectures ($3^o$ cycl de la Suisse Romande} {1992)}

\ref{K.Osterwalder and R.Stora (eds),}{ Critical Phenomena, Random Fields,
Gauge Theoric
Les Houches (1984),}{ (North Holland, Amsterdam 1986)}

\ref{K.Gawedzki and A.Kupiainen:}{}{ Nucl. Phys. B 262, (1985) 33, Comm. math.
Phys. {\bf
102}, 1985, 1}

\ref{K.Gawedzki and A.Kupiainen:}{}{ Comm. Math, Phys, {\bf 89}, 191 (1983)}

\end